\pdfoutput=1

\documentclass[8.5pt,twoside,twocolumn]{article}
\usepackage[super,sort&compress,comma]{natbib}
\usepackage{mhchem}
\usepackage{times,mathptm}
\usepackage{sectsty}
\usepackage{balance}

\usepackage{graphicx} 
\usepackage{lastpage}
\usepackage[format=plain,justification=raggedright,singlelinecheck=false,font=small,labelfont=bf,labelsep=space]{caption}
\usepackage{fancyhdr}
\begin{document}

\twocolumn[
 \begin{@twocolumnfalse}
\noindent\LARGE{\textbf{Metal-adsorbed graphene nanoribbons}}
\vspace{0.6cm}

\noindent\large{\textbf{S. Y. Lin,\textit{$^{a}$} M. F. Lin\textit{$^{b\ast}$}}}\vspace{0.5cm}

\noindent\textit{\small{\textbf{Received Xth XXXXXXXXXX 20XX,
Accepted Xth XXXXXXXXX 20XX\newline First published on the web Xth
XXXXXXXXXX 200X}}}

\noindent \textbf{\small{DOI: 10.1039/b000000x}}
\vspace{0.6cm}

\noindent \normalsize{The metal atoms, the alkali ones excepted, might provide the multiple outermost orbitals for the multi-orbital hybridizations with the out-of-plane $\pi$ bondings on the honeycomb lattice. This will dominate the fundamental properties of Al-, Ti- and Bi-adsorbed graphene nanoribbons, in which they are explored thoroughly by using the first-principles calculations. The principle focuses are the adatom-dependent binding energies, the adatom-carbon lengths, the optimal position, the maximum adatom concentrations, the free electron density transferred per adatom, the adatom-related valence and conduction bands, the various van Hove singularities in DOSs, the transition-metal-induced magnetic properties, and the significant competitions of the zigzag edge carbons and the metal/transition metal adatoms in spin configurations. The distinct chemical bondings are clearly identified from three kinds of metal adatoms under the delicate physical quantities. The important differences between Al-/Ti-/Bi- and alkali-adsorbed graphene nanoribbons will be discussed in detail, covering band structures, relation of conduction electron density and adatom concentration, spatial charge distributions, orbital-decomposed DOSs, and magnetic configurations $\&$ moments.} \vspace{0.5cm}
\end{@twocolumnfalse}
  ]
\

\section{INTRODUCTION}
\footnotetext{\textit{$^{a}$~Department of Physics, University of Houston, TX 77004, USA.}}
\footnotetext{\textit{$^{b}$~Department of Physics, National Cheng Kung University, 701 Tainan, Taiwan.}}
\footnotetext{$\ast~$ E-mail: mflin@mail.ncku.edu.tw}

Up to now. there are a lot of studies associated with aluminium-based batteries. On the graphite cathode,\cite{lin2015ultrafast} the predominant AlCl$_4$ molecules are,respectively, intercalated and de-intercalated between graphitic layers during charge and discharge processes. On the aluminium anode, the metallic Al and AlCl$_4$ are transformed into Al$_2$Cl${_7^{-}}$ during discharging, and the reverse chemical reactions happen under charging. Al atoms are deduced to play a critical role in the giant enhancement of current densities. The substitution of lithium with aluminum anode, which is characterized by the lower cost and higher abundance, is one of the most widely studied methods to reduce the cost of electrochemical storage systems and enable the long-term sustainability.\cite{elia2016overview} On the theoretical side, the first-principles calculations on Al-doped graphenes predict that such systems are suitable in applying to environment and energy engineerings, such as toxic gases senors/detectors\cite{ao2008enhancement,chi2009adsorption} and a potential hydrogen storage.\cite{ao2009doped} Moreover, the Al-absorbed 2D monolayer graphene shows the maximum adatom concentration corresponding to a ${2\times\,2}$ enlarged unit cell.\cite{lin2017feature} The optimal position is the hollow site with the height of ${2.00}$ $\AA$ and the Al-C bond length of ${2.54}$ $\AA$. This system exhibits the red shift of the Fermi level, in which the well-known Dirac cone, with the linear energy dispersions, is preserved or only slightly distorted. That is to say, the low-lying energy bands are dominated by the original $\pi$ bonding of carbons. The free electron densities arising from the $n$-type dopings are very high and quite close to those observed in alkali-adsorbed graphenes. There no magnetic properties after aluminum adsorption. However, there exist the significant multi-orbital hybridizations in Al-C bonds, namely, the ${(3s,3p_x,3p_y)}$ and ${2p_z}$ orbitals. The main reason might be that each Al adatom has three outmost valence electrons. They are revealed as the (Al,C)-co-induced valence and conduction bands out of the Dirac-cone structure and the merged peaks in the orbital-decomposed DOSs. The finite-size effects and the edge structures are expected to greatly modify the electronic and magnetic properties.

In recent years, various doping methods for epitaxial few-layer graphenes are realized by adatom adsorption and substitution. The previous experimental and theoretical study on the titanium-absorbed graphene has confirmed the greatly modified band structures by using the high-resolution ARPES measurements.\cite{chen2016efficient} By depositing a low concentration of Ti atoms on epitaxial graphene supported by SiC, the electronic band structure of graphene could be significantly engineered.
The titanium low coverage is explored in experimental measurements, since that Ti adatoms are easy to from nanoclusters rather than isolated adatoms.\cite{chan2008first,mashoff2013hydrogen} Such adatoms on graphene surface are predicted to have large cohesive energies, indicating there exists the critical orbital hybridization in the Ti-C bonds. The first-principles calculations reveal that a very strong hybridization occurs primarily between the vertical Ti-3$d$ and C-2${p_z}$ orbitals as a result of the spatial overlap and symmetry matching. This is responsible for the high doping efficiency and the readily tunable carrier density. This Ti-induced hybridization is adsorbate-specific and creates major consequences for the efficient doping as well as the potential applications towards adsorbate-induced modification of carrier transport in graphene. For example, the titanium-covered graphene is proved to be very efficient in hydrogen adsorption and desorption,\cite{mashoff2013hydrogen} but not the formation of H$_2$ molecules.

Nowadays, the bismuth-related systems are one of the most widely studied materials, because the dimensionality could enrich the fundamental properties. The bulk bismuth, which possesses a rhombohedral symmetry, is a semimetal with a long Fermi wavelength and small effective electron mass.\cite{hofmann2006surfaces} Its surface states belong to a Dirac fermion gas.\cite{li2008phase} The three bismuth surfaces: Bi(111), Bi(100), and Bi(110) have a higher free carrier density at the Fermi level compared to those of the bulk system.\cite{hofmann2006surfaces} The Bi thin film, without strong chemical reactions by O$_2$ on surfaces, appears to be stable up to about 600 K.\cite{bobaru2012competing} Recently, the 2D few-layer bismuthenes are epitaxially grown the 3D Bi$_2$Te$_3$(111)/Bi$_2$Se$_3$(111)/Si(111) substrates.\cite{hirahara2006role,hirahara2012atomic,yang2012spatial,wang2013creation,hirahara2015role} Also they have been successfully obtained from the mechanical exfoliation.\cite{sabater2013topologically} Monolayer bismuthene is predicted to exhibit the rich and unique magnet-electronic properties, mainly owing to the cooperation among the specific geometric symmetry, the multi-orbital hybridizations in Bi-Bi bonds, the significant spin-orbital couplings, the magnetic field and the electric field.\cite{chen2017novel} Moreover, the 1D bismuth nanowires exhibit narrow band gaps due to significant quantum confinement effects.\cite{black2002infrared} The above-mentioned 3D$-$1D bismuth systems have the significant ${sp^3}$ bondings and the strong spin-orbital couplings, in which the former mainly come from the ${(6s,6p_x,6p_y,6p_z)}$ orbitals of Bi atoms. The ${sp^3}$ tight-binding models, with the spin-orbital interactions, are successful in understanding the magneto-electronic properties of the semi-metallic 3D rhombohedral bismuth\cite{PhysRevB.52.1566} and the semi-conducting 2D bismuthene.\cite{Jason2015Two} In addition, bismuth systems are explored in detail for the fields of environmental engineering, biochemistry, and energy engineering, such as, Bi-based nanoelectrode arrays in detecting heavy metals,\cite{wanekaya2011applications} a polycrystalline bismuth oxide film as a biosensor,\cite{shan2009polycrystalline} and bismuth oxide on nickel foam available for the anode of lithium battery.\cite{li2013bismuth}

Recently, bismuth adatoms on monolayer graphene supported by a 4H-SiC(0001) substrate have been obviously observed at room temperature.\cite{chen2015long,chen2015tailoring} The corrugated substrate and buffer graphene layer are clearly identified from the STM experiments, in which a large-scale hexagonal array of Bi adatoms appears at room temperature. Such adatoms could form the triangular and rectangular nanoclusters of a uniform size by the further annealing process. Moreover, the STS measurements of the dI/dV spectra confirm the existence of the Dirac-cone structure by the $V$-shape differential conductance, indicating the blue shift of the Fermi level (the $p$-type doping) and the creation of free conduction electrons. In addition, the Bi-related structures in DOS are revealed at deeper energy. These important results shed light on controlling the nucleations of the isolated adatoms and nanostructures on graphene surfaces. They are successfully simulated by the six-layered substrate, the corrugated buffer layer, and the slightly deformed monolayer graphene using the first-rinciples calculations.\cite{lin2016substrate} The Bi adatom arrangements are thoroughly invetigated by analyzing the ground state energies, bismuth adsorption energies, and Bi¡VBi interactions of energies under the different heights, inter-adatom distances, adsorption sites, and hexagonal positions. A hexagonal array of Bi adatoms is dominated by the van der Waals interactions between the buffer layer and monolayer graphene. An increase in temperature could overcome a ${\sim\,50}$ meV energy barrier and generate the unusual triangular and rectangular nanoclusters. The most stable and metastable structures are consistent with the experimental observations. There also exist some theoretical studies on the geometric structures and energy bands of Bi-adsorbed and Bi-intercalated graphenes.\cite{akturk2010bismuth,hsu2013first} The former is conducted on monolayer graphene without simulation of the substrate and the buffer graphene layer, and thus the deformed graphene surface structure might be unreliable.\cite{akturk2010bismuth} The latter is evaluated for Bi and/or Sb as a buffer layer above the four-layer SiC substrate, which lead to an energetically unfavorable environment for the metal adatoms to be adsorbed on the graphene sheet.\cite{hsu2013first}

Metal atoms, including aluminum, titanium and bismuth, could provide effective dopings in conduction electrons (the $n$-type dopings),\cite{lin2017feature,chen2016efficient,chen2015long,chen2015tailoring} as revealed in alkali ones. After adsorption on graphene naoribbon surfaces, they are expected to induce more complicated multi-orbital hybridizations in the significant carbon-metal bonds, compared with the ${2p_z}$ orbital and the outermost ${s}$ orbital in the carbon-alkali bonds. Based on detailed first-principles calculations on the Al-, Ti-, and Bi-adsorbed graphenes, the critical bondings are identified to arise from the ($3s,3p_x,3p_y$), ($3d_{z^2},3d_{xy},3d_{x^2-y^2}$) and the ($6s,6p_x,6p_y,6p_z$) orbital, respectively. The theoretical predictions could highly promote further understanding of experimental measurements. For example, the aluminum-based batteries have been developed quickly to greatly enhance the charging and discharging reactions and to reduce the cost of the metallic anode.\cite{lin2015ultrafast} The Al-adsorbed graphene is predicted to create as many free carriers as the alkali adatoms do, while the concentration of the former has an upper limit of ${25\%}$.\cite{lin2017feature} When the low-coverage titanium adatoms are adsorbed on monolayer graphene supported by 4H-SiC(0001) substrate, the high-density free electrons are verified from the high-resolution ARPES measurements.\cite{chen2016efficient} The theoretical calculations suggest that the Ti adsorptions could exhibit the high-concentration adsorptions. The measured results agree with the calculated ones under the very strong orbital hybridizations in the Ti-C bonds. Specially, a large-scale hexagonal array of Bi adatoms on graphene surface is clearly revealed at room temperature using the STM measurements.\cite{chen2015long,chen2015tailoring} It becomes three- and four-member nanostructures under the annealing process. The six-layer SiC substrate, the buffer graphene layer, and the slightly deformed monolayer graphene in the first-principles model are proposed to explain the most and meta-stable optimal structures.\cite{lin2016substrate} The critical orbital hybridizations in generating free conduction electrons are also examined in detail.\cite{lin2016substrate} The limit of adatom concentration, the metal-induced free carrier density, and the magnetic properties are focuses of the metal-adsorbed graphene nanoribbons.

\section{COMPUTATIONAL DETAILS}
The geometric and electronic structures of metal-adsorbed graphene nanoribbons are studied by the Vienna ab initio simulation package\cite{kresse1999ultrasoft,kresse1996efficient} in the density-functional theory (DFT). The DFT-D2 method\cite{grimme2006semiempirical} is taken into accountin order to describe the weak van der Waals interactions.The projector augmented wave method is utilized to characterize the electron-ion interactions. The exchange-correlation energy of the electron-electron interactions is evaluated within the local-density approximation. The wave functions are expanded by plane waves with the maximum kinetic energy limited to $500$ eV. The k-point sampling is outlined by the Monkhorst-Pack scheme.\cite{perdew1996generalized} The $12\times1\times1$ and $300\times1\times1$ k-grids in the first Brillouin zone are, respectively, the settings used for the geometry optimization and band-structure calculations. The Hellmann-Feynman net force on each atom is smaller than $0.03$ eV/{\AA}. In order to avoid interactions between the scrolled graphene superlattices of the adjacent unit cells, different vacuum spacings in the z-direction (or the y-direction) are tested and a value of $15$ {\AA} is best for the accurate and efficient calculations.

\section{Results and discussion}
\subsection{Al}

\begin{figure}[h]
\centering
  \includegraphics[width=1.0\linewidth]{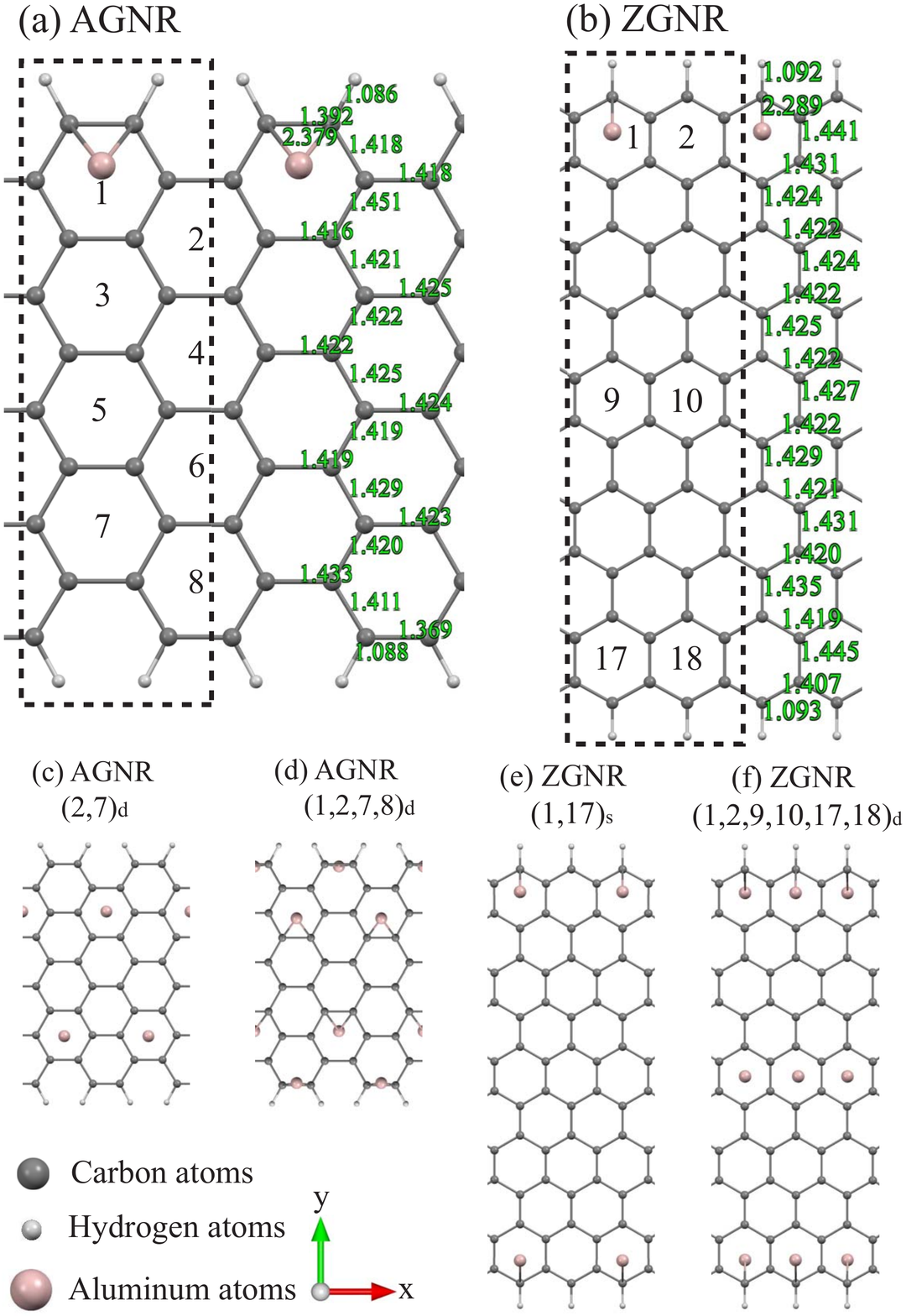}
  \caption{The non-uniform optimal geometric structures for the Al-adsorbed  ${N_A=10}$ armchair graphene nanoribbons under the initial adatom positions: (a) (1)$_s$ (c) (2,7)$_s$ $\&$ (d) (1,2,7,8)$_d$; those of  the ${N_Z=10}$ systems with the similar positions: (b) (1)$_s$, (e) (1,17)$_d$ $\&$ (f) (1,2,9,10,17,18)$_d$}
  \label{fgr:1}
\end{figure}

The Al-adsorbed graphene nanoribbon remains the planar structure with a non-uniform honeycomb lattice, as revealed in Figs. 1(a)-1(f). This clearly indicates the well-behaved $\sigma$ bondings of carbon atoms, being hardly affected by the Al adsorptions. The optimal positions might correspond to the hollow sites if the adatoms are far away from the boundaries, i.e., the ${(x,y)}$-plane projections are the centers of hexagon lattice in the absence of $y$-shift (Table 1). But when the Al adatoms are located near the armchair and zigzag boundaries (Figs.1(a)-1(f)), the obvious shifts are revealed along the transverse direction toward the edges, e. g., ${\sim\,0.25-0.32}$ $\AA$ for the 1, 2, 7 $\&$ 8 positions in ${N_A=10}$ armchair system and ${\sim\,0.06-0.17}$ $\AA$ for the 1, 3 $\&$ 5 positions in ${N_z=10}$ zigzag one. The height of Al adatom is ${2.07-2.21/2.00-2.14}$ $\AA$ for armchair/zigzag systems, in which the maximum value is associated with the nanoribbon center. That is, the Al adatoms are relatively low under the effect of the edge C-H bonds. As to the highest concentration, the stable structure is associated with the double-side adsorption (not the single-side adsorption) of four Al adatoms in ${N_A=10}$ armchair system, and it is related to the similar adsorption of seven Al adatoms in ${N_z=10}$ zigzag one. This result does not exceed the upper limit of 25${\%}$ in Al-adsorbed monolayer graphene.\cite{lin2017feature} There exist the unusual geometric structures under the maximum concentration, as obviously indicated in Figs. 1(d) and 1(f). The optimal position is dramatically transferred to the bridge side, clearly illustrating the complex competitions/cooperations among the C-C, Al-C, Al-Al and H-C chemical bondings. This means that the two Al adatoms need to have the sufficiently long distance to achieve the stable structure. The high-resolution STM and TEM could be utilized to verify the first-principles predictions. After Al adsorption, the similar 1D graphene plane suggests that only the ${2p_z}$ orbitals of carbons have significant chemical bondings with the half occupied three kinds of orbitals (${3s,3p_x,3p_y}$), and the $\sigma$ bondings of ${(2s,2p_x,2p_y)}$ almost keep the same. This will be explored thoroughly.

\begin{figure}[h]
\centering
  \includegraphics[width=1.0\linewidth]{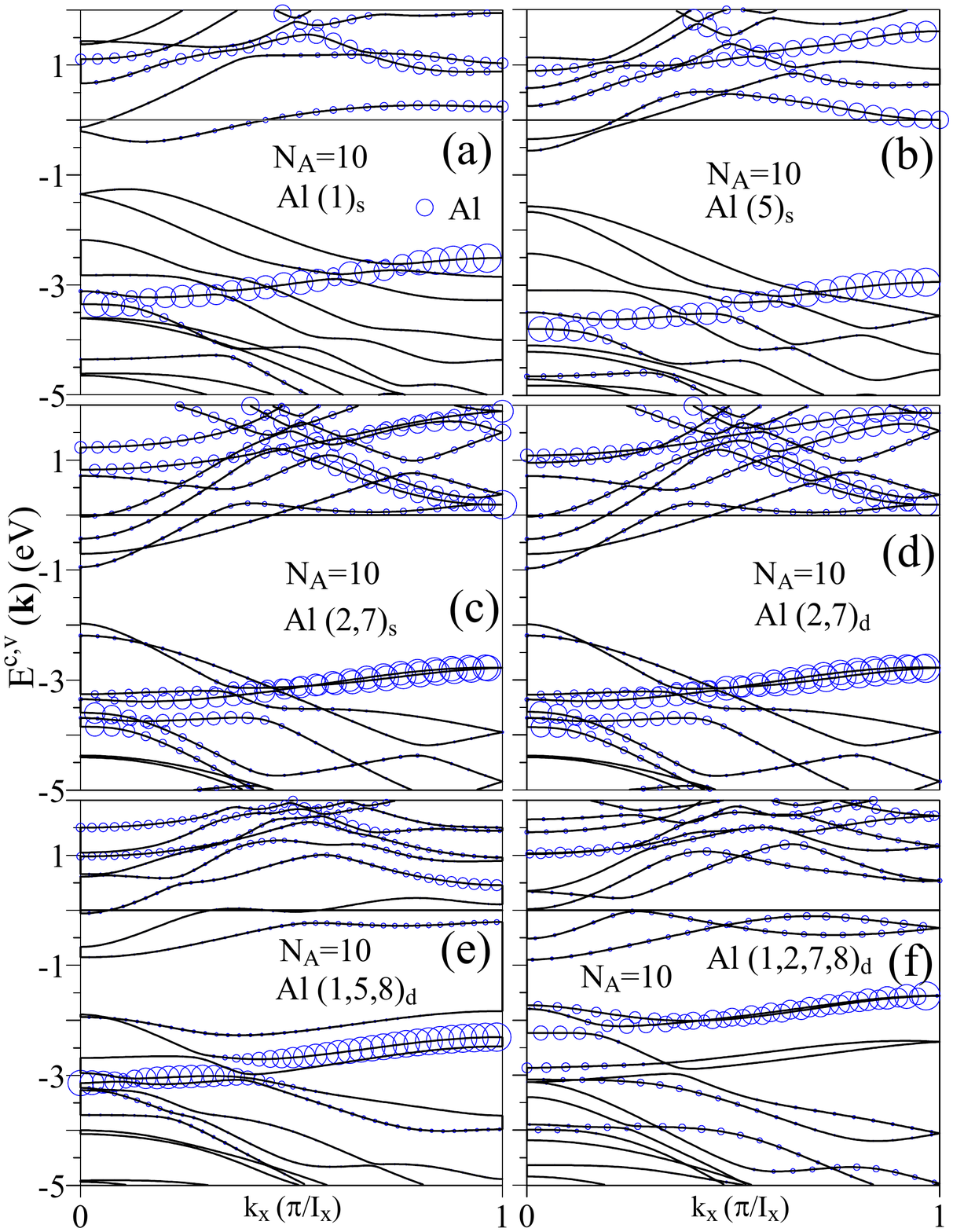}
  \caption{Band structures of the Al-adsorbed ${N_A=10}$ armchair graphene nanoribbons for the various adsorptions: (a) (1)$_s$, (b) (5)$_s$, (c)  (2,7)$_s$, (d) (2,7)$_d$, (e) (1,5,8)$_d$  and  (f) (1,2,7,8)$_d$.}
  \label{fgr:1}
\end{figure}

The electronic structures exhibit the drastic changes in all graphene nanoribbons during the aluminization, as clearly shown in Figs. 2 \& 3. In general, the Fermi level is shifted from the center of energy gap to the conduction bands. Furthermore, the energy bands very close to $E_F$ are mainly determined by carbon atoms, indicating the free electrons due to the distorted $\pi$ bondings and the charge transfer from carbon atoms to Al adatoms. For armchair systems, the energy spacing between the first pair of valence and conduction bands is related to ${k_x\neq\,0}$ or ${k_x=0}$, in which the former might depend on whether the Al adatoms are situated at armchair edges (the (1)$_s$ case in Fig. 2(a)). Specifically, the Al adatoms make important contributions to certain conduction and valence bands in the range of ${0\le\,E^c\le\,2.0}$ eV and ${-2.5}$ eV${\le\,E^v\le\,-4.3}$ eV, in which the ${-2.5}$ eV${\le\,E^v\le\,-3.5}$ eV valence states are weakly dispersive and almost doubly degenerate. The Al adsorption could create the extra band-edge states arising from the subband hybridizations. Energy bands are very sensitive to the changes in the position and concentration of adatoms (Figs. 2(a)-2(d)), but not the single- or double-side adsorption (Figs. 2(c) and 2(d)).

\begin{figure}[h]
\centering
  \includegraphics[width=1.0\linewidth]{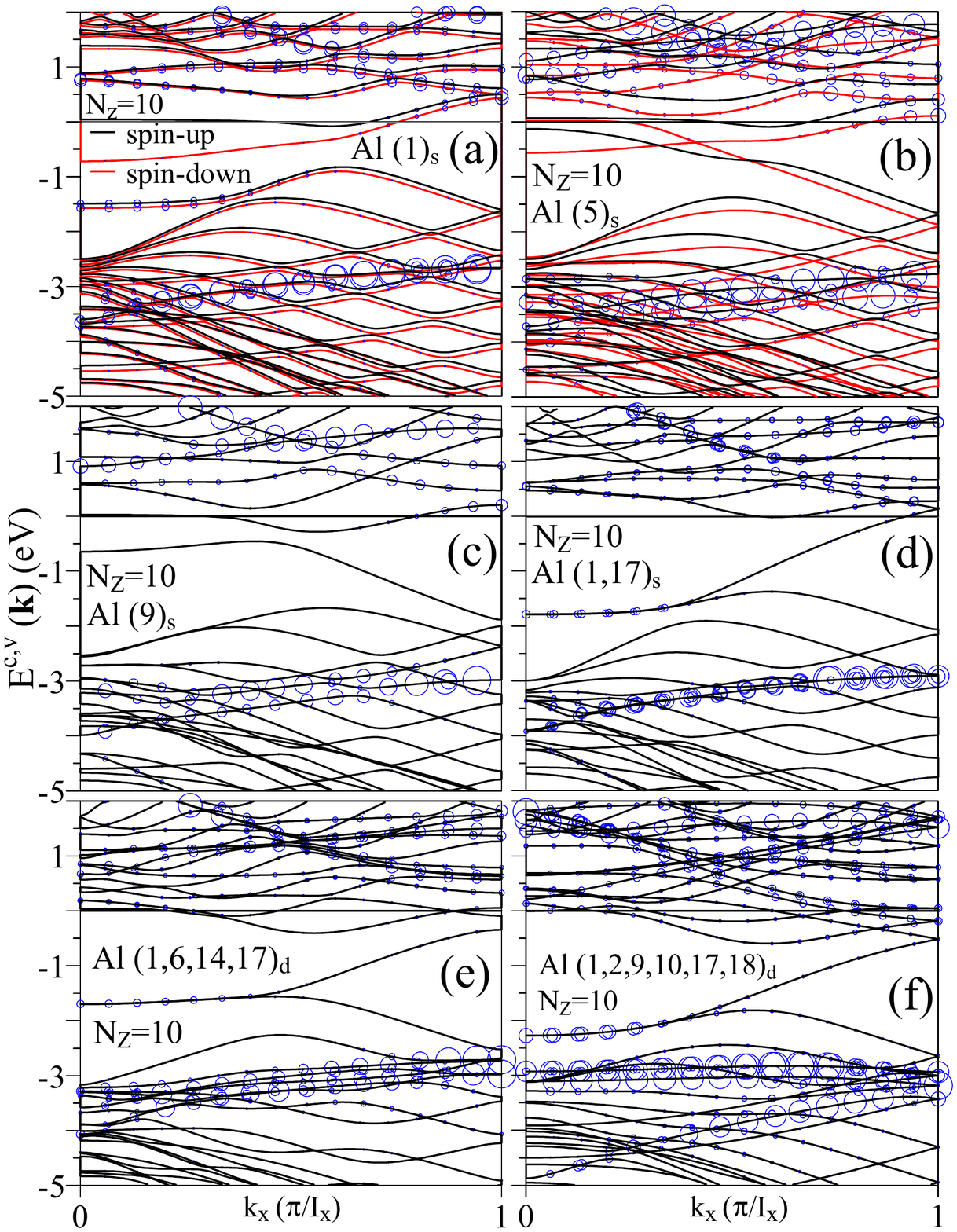}
  \caption{Similar plot as Fig. 2, but displayed for the ${N_Z=10}$ zigzag systems under the various cases: (a) (1)$_s$, (b) (5)$_s$, (c) (9)$_s$, (d) (1,17)$_s$, (e) (1,6,14,17)$_d$  and  (f) (1,2,9,10,17,18)$_d$.}
  \label{fgr:1}
\end{figure}

The free conduction electrons provided by per Al adatom in a unit cell are worthy of closer examinations. With the increasing Al concentration, there are more carbon-dominated conduction bands intersecting with the Fermi level, as obviously revealed in Figs. 2(c)-2(d) for aluminization with two adatoms. By the addition of the Fermi momenta, the 1D linear carrier density becomes double, regardless of the single- or double-side adsorption, and the adatom positions. The linear relation between conduction electron density and adatom concentration is roughly achieved under the specific three Al adsorptions, e.g., the ${(1,5,8)}$ and ${(1,2,8)}$ cases. However, this simple relation is absent under the maximum adsorption concentration. The four-Al case might correspond to a indirect-gap semiconductor with a very small $E_g$, e.g, ${E_g=0.04}$ eV under the (1,2,7,8)$_d$ adsorption.

Energy bands strongly depend on the edge structures; that is. there are certain important differences in zigzag and armchair graphene nanoribbons. For the single adatom adsorption on a zigzag system, whether the low-lying energy bands exhibit the spin spitting is very sensitive to the position of Al, as clearly indicated in Figs. 3(a)-3(d). There exist the drastic changes in energy bands, especially for electronic states near ${k_x=0}$ and ${\pm\,2/3}$. A pair of partially flat and spin-degenerate valence and conduction bands, which, respectively, cross the Fermi level and are almost symmetric about ${E_F=0}$, dramtically change the energy dispersions, possess the Fermi momenta in the conduction bands, and even present the spin spitting. When the Al adatom is not located at the nanoribbon center, the spin-spilt metallic behavior comes to exist, as shown in Figs. 3(a)-3(c). Specially, the spin-split energies arising from the partially flat bands even reach ${\sim 0.5-0.7}$ eV close to the ${k_x=0}$ state. The occupied state number is different in the spin-up- and spin-down-related energy bands; that is, such zigzag systems belong to the FM metal, in which the net magnetic moment is 0.56 $\mu_B$. per unit cell. On the other hand, the central single-Al adsorption cannot create the spin splitting and thus preserve the original AFM spin configuration ((the (9)$_s$ case in Fig. 3(d)). Band structures become more complicated in the increase of adatom concentration. When there are two Al adatoms close to the distinct boundaries, the magnetic configuration are fully absent. For example, the ${(1.17)}$ adsorption in Fig. 3(e) is a non-magnetic metal, in which the valence and conduction energy bands are doubly degenerate near the ${k_x=0}$ state because of the merged neighboring bands.

\begin{figure}[h]
\centering
  \includegraphics[width=1.0\linewidth]{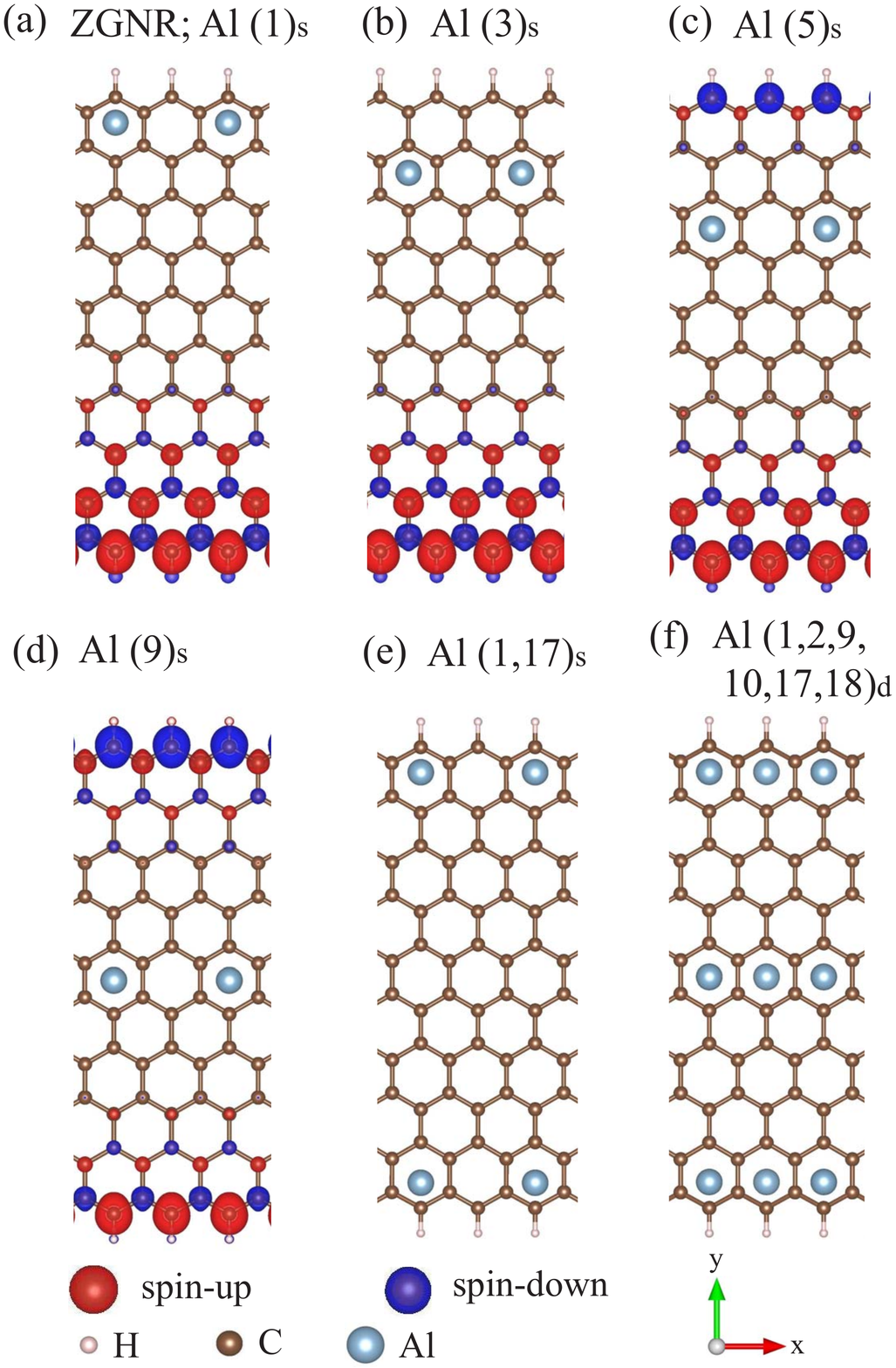}
  \caption{The spatial spin distributions of the Bi-adsorbed ${N_Z=10}$ zigzag systems due to the various adsorptions: (a) (1)$_s$, (b) (3)$_s$,  (c) (5)$_s$, (d) (9)$_s$, (e) (1,17)$_d$, and (f) ${(1,2,9,10,17,18)_d}$. They are shown on the ${x-y}$ plane.}
  \label{fgr:1}
\end{figure}

The Al-adsorbed zigzag graphene nanoribbons exhibit the diverse magnetic configurations, being different from/similar to the spin arrangements of the alkali-adsorbed ones. A single Al close to the upper zigzag boundary will partially destroy the spin configutation due to the zigzag edge carbons, depending on the adatom positions, as clearly revealed in Figs. 4(a)-4(c). The lower zigzag edge preserves the spin-up-dominated configuration (red balls) almost identical to the original case. However, the spin-down arrangement (blue balls) near the upper edge is fully/partially annihilated under the (1)$_s$ $\&$ (3)$_s$/(5)$_s$ cases, and the spin-up distribution might be drastically changed from edge to center. Specially, the spin-up configuration in the whole zigzag nanoribbon has been created by the (1)$_s$ and (3)$_s$ adsorptions (Figs. 4(a) and 4(b)); that is, the edge and non-edge carbons possess the magnetic configurations simultaneously. This is absent in the alkali-adsorption systems. It should have the maximum magnetic moment for the Al adatom near the zigzag edge. On the other hand, one central Al adatom hardly affects the original AFM spin distribution, being similar to the alkali case. The magnetic properties thorough vanish under two Al adatoms near the separate zigzag edges, as observed in alkali-adsorbed systems. The above-mentioned features of magnetic properties clearly show the very strong competition/cooperation between the multi-orbital hybridizations of Al-C bonds and the spin sates due to zigzag edge carbons in the total ground state energy.

\begin{figure}[h]
\centering
  \includegraphics[width=1.0\linewidth]{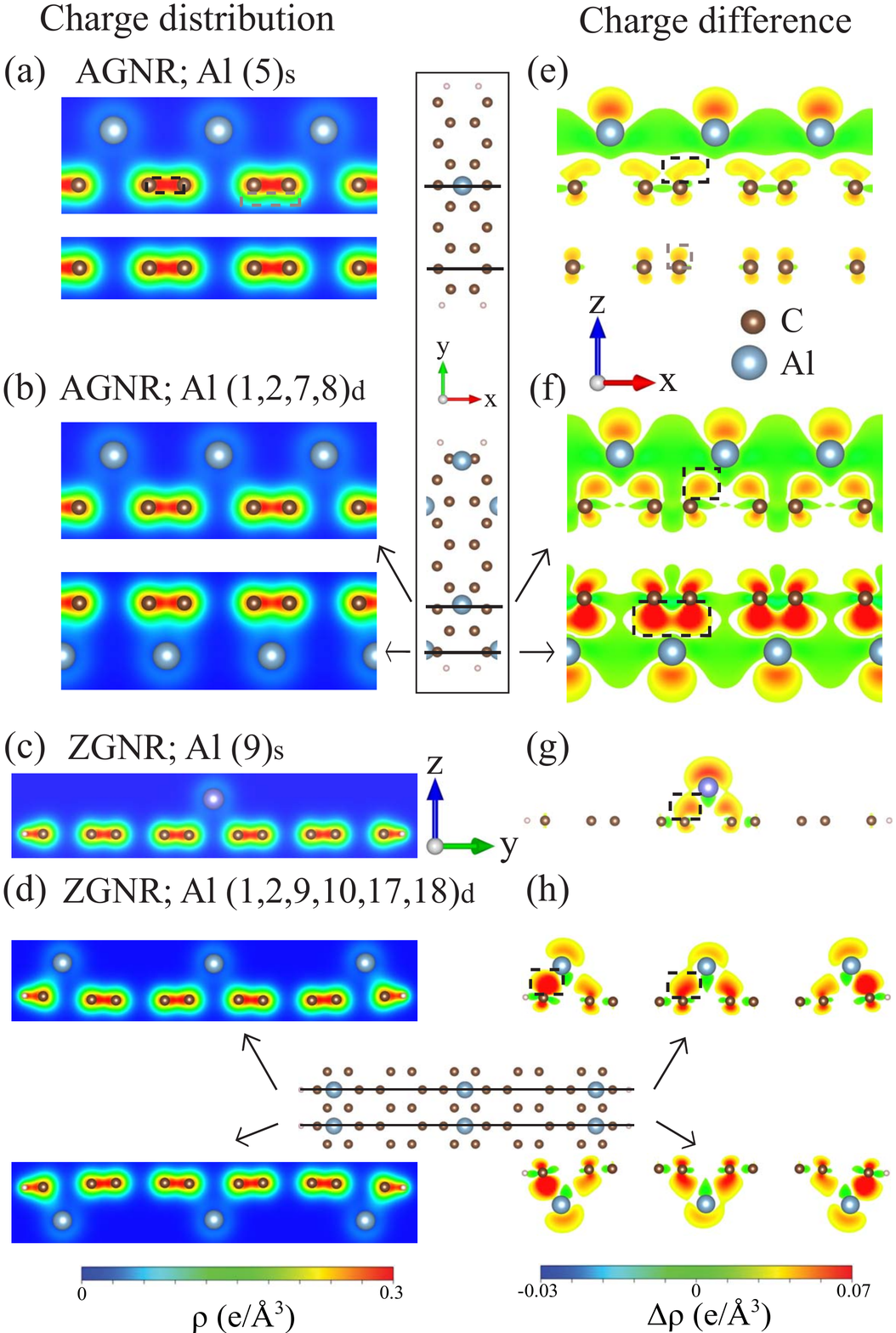}
  \caption{The spatial charge distributions and the differences after chemisorptions. $\rho$'s/${\Delta \rho}$ on the ${x-z}$ plane are shown for the ${N_A=10}$ armchair systems with Bi adatoms at (a)/(e) (5)$_s$ $\&$ (b)/(f) (1,2,7,8)$_d$; those on the ${y-z}$ plane for the ${N_Z=10}$ zigzag systems under the Bi-adsorptions: (c)/(g) (9)$_s$ $\&$ (d)/(h) (1,2,9,10,17,18)$_d$.}
  \label{fgr:1}
\end{figure}

The carrier density and the variation of carrier density can provide very useful informations about the orbital bondings, energy bands, and charge transfer. The former directly reveals the bonding strength of C-C, Al-C, and Al-Al bonds, as illustrated in Figs. 5(a)-5(d). After the Al adatom adsorptions, all the C-C bonds possess the strong covalent $\sigma$ bonds (blue rectangle) and the somewhat weaken $\pi$ bonds simultaneously (red rectangle). The former almost keep the same; furthermore, the $\pi$ bonding also belongs to the extended state in a 1D system except that it is seriously distorted under the maximum-concentration case (Fig. 2(f)). These are responsible for the carbon-dominated low-lying energy bands with the metallic or semiconducting behavior (Fig. 2). In general, the ${3s}$-orbital electrons (black triangle) are redistributed between Al and the six nearest C atoms, revealing a significant hybridization with the ${2p_z}$ orbitals (yellow rectangle; green ring in the inset). The spatial charge distribution of (${3p_x,3p_y)}$) orbitals is extended from the light blue ring (inset) near Al to the red ring between C and Al atoms. It is noticed that these orbitals make less contribution to the chemical bonding. The multi-orbital hybridizations of ${3s}$ and ${(3p_x,3p-y)}$ are, respectively, related to the Al-dominated valence and conduction bands (Fig. 2). The electrons of Al adatom move to the top and bottom of the non-nearest C atoms (red region within a black rectangle), inducing free electrons in conduction bands. The higher the Al-concentration is, the more the electrons are transferred (Fig. 4(f)).

\begin{figure}[h]
\centering
  \includegraphics[width=1.0\linewidth]{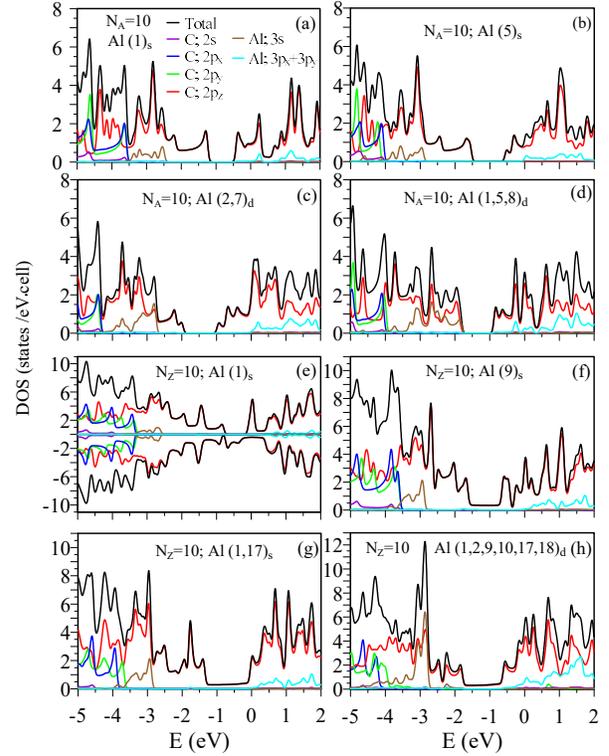}
  \caption{The orbital- and spin-projected DOSs for the ${N_A=10}$ armchair systems under the various adsorptions: (a) (1)$_s$, (b) (5)$_s$, (c) (2,7)$_d$; (d)   (1,5,8)$_d$, and those for the ${N_Z=10}$ systems  with the adatom positions: (e) (1)$_s$, (f) (9)$_s$, (g) (1,17)$_s$, $\&$ (h) (1,2,9,10,17,18)$_d$.}
  \label{fgr:1}
\end{figure}

The Al-adatom adsorptions can create the metallic DOSs in graphene nanoribbons except for the maximum concentration (Figs. 6(a)-6(h)), while the main characteristics of van Hove singularities are in sharp contrast with those of H-terminated pristine systems and alkali-adsorbed cases. The low-energy DOSs are dominated by the ${2p_z}$ orbitals of carbons, being consistent with the atom-dominated band structures (Figs. 2 and 3). This feature clearly illustrates that the $\pi$ bondings are distorted only near the Al adatoms, and they behave as the normal extended states at others. For armchair systems (Figs. 6(a)-6(d)), there exist the specific zero DOSs below the Fermi level in the range of ${-2}$ eV${\le\,E^v\le\,-1}$ eV, belonging to the initial valence and conduction bands. However, the similar DOSs have a finite value in zigzag systems (Figs. 6(e)-6(h)). Concerning the clear evidences of the multi-orbital hybridizations in Al-C bonds, they show as the merged peaks in the range of ${E^v<-1.5}$ from the Al-${3s}$ orbital and the C-${2p_z}$ orbital, and those for ${E^c>-0.5}$ eV due to the Al-(3p$_x$,3p$_y$) orbitals and the C-${2p_z}$ orbital. Energy bandwidths of the ${3s}$ and ${(3p_x+3p_y)}$ orbitals grows gradually in the increment of Al concentrations, indicating two kinds of orbital hybridizations in Al-Al bonds. It should be noticed that the conduction states below the Fermi level about ${\sim\,1}$ eV also make inportnt contributions to the essential properties. The ${3p_z}$ orbitals of Al adatoms hardly contribute to the significant chemical adsorption, since each adatom only has three occupied orbitals in the outermost ones, in which it  might possess two states in the ${3s}$ orbitals and one state in the ${(3p_x,3p_y)}$ orbitals  (from the covered areas by the ${3s}$ and ${3p_x+3p_y}$ orbitals, respectively). Specifically, it might be difficult to identify the special peaks coming from the low-lying partially flat bands, and the spin-up- and spin-down-split DOSs are obviously revealed in certain adsoption configurations of zigzag graphene nanoribbons (Figs, 6(e) and 6(h)). The FM metals are useful in exploring the spintronic transports and the spin-related applications.

\subsection{Ti}

\begin{figure}[h]
\centering
  \includegraphics[width=1.0\linewidth]{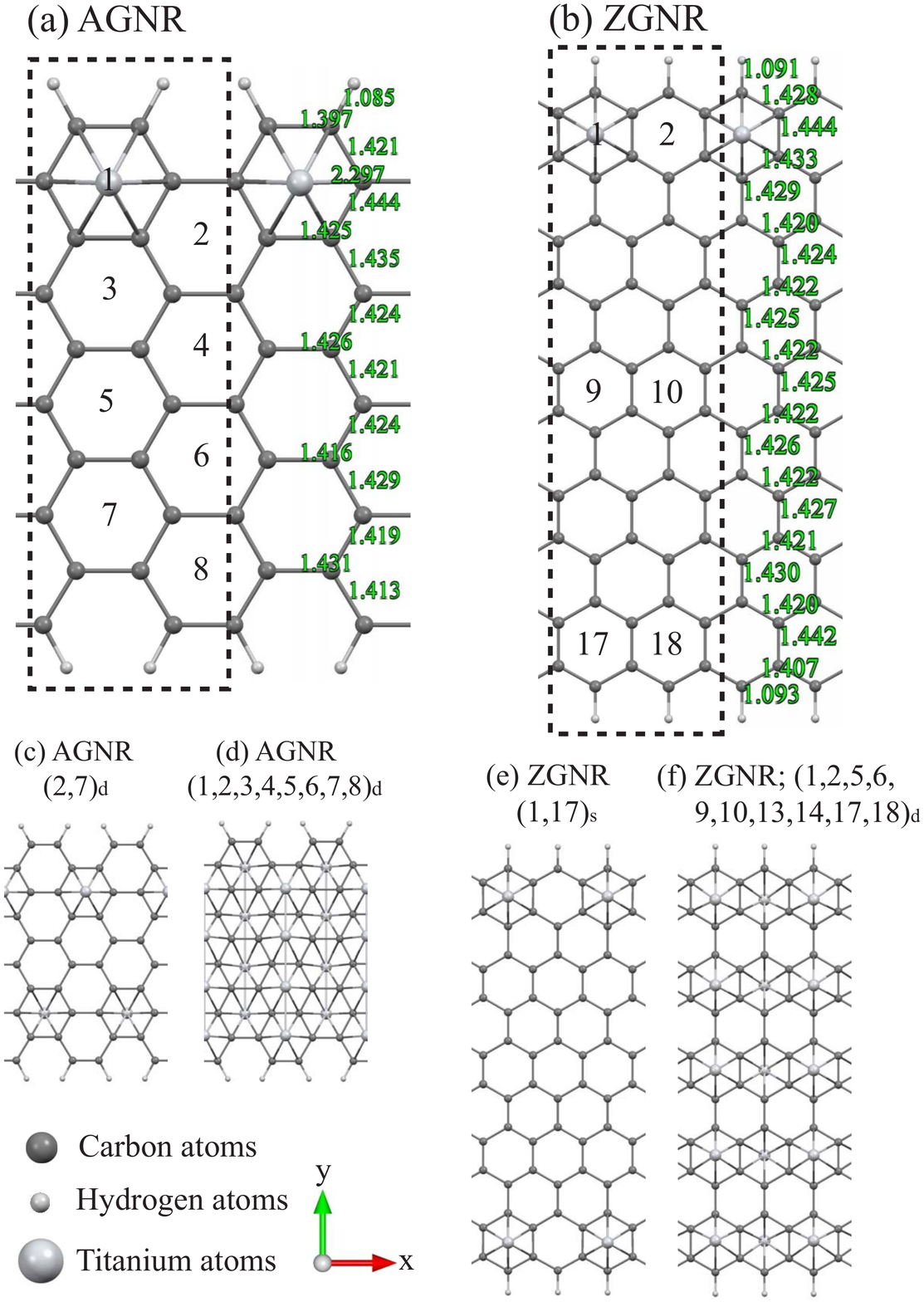}
  \caption{The optimal geometric structures for the Ti-adsorbed  ${N_A=10}$ armchair systems under the adatom positions: (a) (1)$_s$ (c) (1,8)$_s$ $\&$ (d) (1,2,3,4,5,6,7,8)$_d$; those of  the ${N_z=10}$ zigzag systems with the similar positions: (b) (1)$_s$, (e) (1,17)$_d$ $\&$ (f) (1,2,5,6,9,10,13,14,17,18)$_d$.}
  \label{fgr:1}
\end{figure}

The Ti-adatom chemisorptions on 1D graphene nanoribbons could create the optimal geometric structures similar to 2D graphene systems except for the non-uniform band lengths. Figures 7(a)-7(f) clearly show the optimal hollow sites without any shifts relative to the ${(x,y)}$-projection centers under all the adsorption configurations. , This is in sharp contrast with the significant shifts of the Al and Bi metal adsorptions (Chaps. 11.1 and 11.3). The Ti heights keep in the short range of ${\sim\,1.50-2.10}$, leading to the largest binding energies (${\sim\,-2.5-3.5}$ eV) among the (Ti,Al,Bi) metal adatoms. It is relatively easy to induce the Ti chemisorptions, compared with the Al and Bi metal adatoms. The former could reach/present the higher adatom concentrations. Specifically, the buckling structures, might be revealed for the highest-concentration adsorption in armcahir/zigzag systems, e.g., (1,2,3,4,5,6,7,8)$_d$/(1,2,5,6,9,10,13,14,17,18)$_d$ in Fig. 7(c)/Fig. 7(f). The main features of optimal structures strongly suggest that the ${2p_z}$ and ${(2s,2p_x,2p_y)}$ orbitals of carbon atoms, respectively, make important and minor contributions to the Ti-C bonds.

\begin{figure}[h]
\centering
  \includegraphics[width=1.0\linewidth]{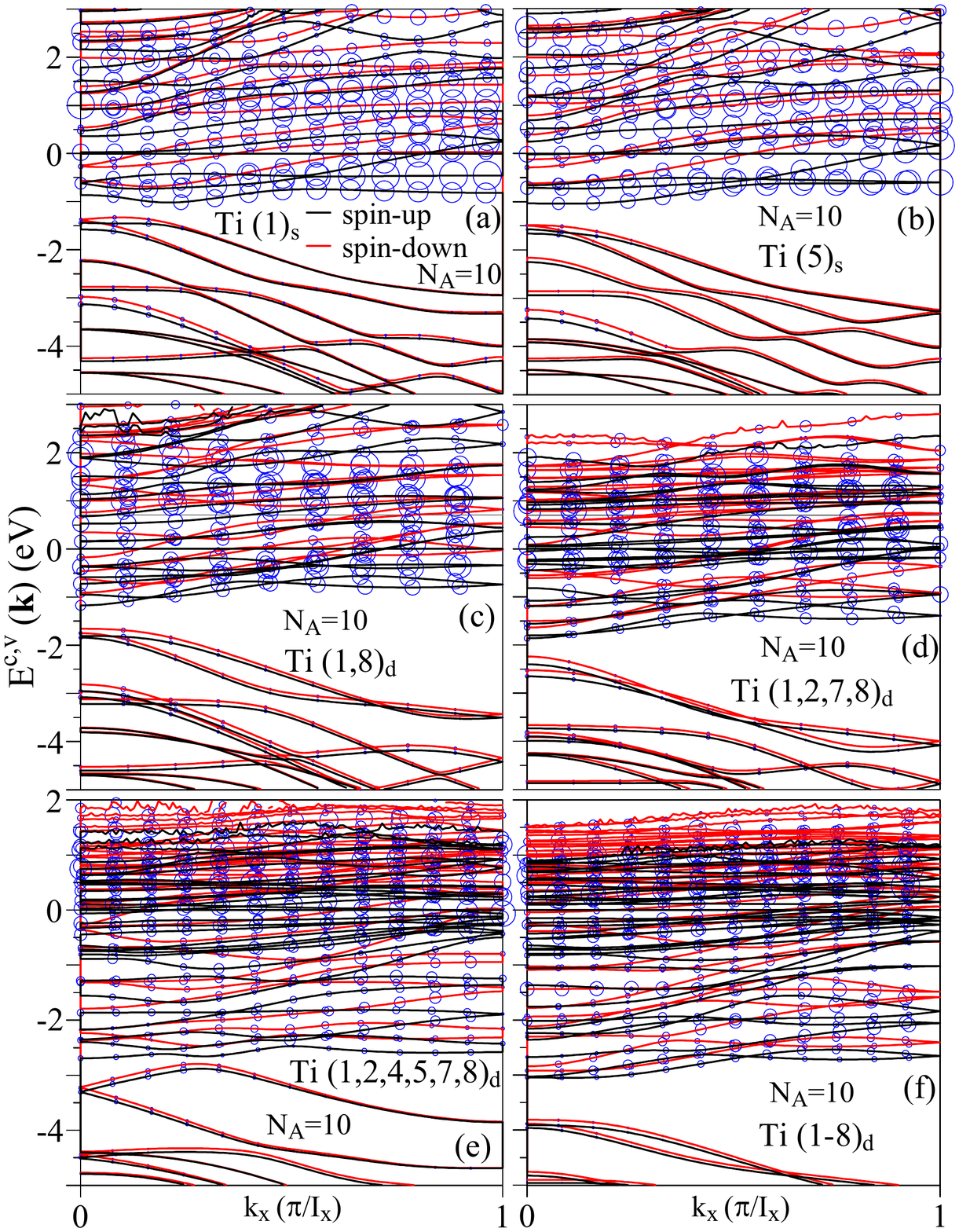}
  \caption{Band structures of Ti-adsorbed ${N_A=10}$ armchair nanoribbons under the various adsorptions: (a) (1)$_s$, (b) (5)$_s$, (c)  (1,8)$_d$, (d) (1,2,7,8)$_d$, (e) (1,2,4,5,7,8)$_d$  and  (f) (1-8)$_d$.}
  \label{fgr:1}
\end{figure}

Any Ti-adsorbed graphene nanoribbons belong to the unusual metals, as clearly indicated in Figs. 8 and 9. There are a lot of energy bands crossing the Fermi level, in which they mainly come from both Ti adatoms and carbon atoms. Furthermore, the Ti-dominated energy bands might be thoroughly occupied or unoccupied. For armchair systems, all of them exhibit the spin-split energy bands, especially for those near $E_F$. Apparently, they correspond to the FM spin configurations, and the net magnetic moments due to the Ti adsorbates are sensitive to the adatom distribution and concentration (Table 1). Under the single- and two-adatom adsorptions (Figs. 8(a)-8(d)), the Ti-dependent electronic structures lie in the range of ${-1}$ eV${\le\,E^{c,v}\le\,3}$ eV, and the host atoms fully determine  the energy bands below it. The increasing adatom concentration, respectively, makes major and minor contributions to these two different ranges of energy bands. The Ti adatoms hardly contribute to the deep valence states, indicating the weak chemical hybridizations from the ${(2s,2p_x,2p_y)}$ orbitals in the Ti-C bonds.

\begin{figure}[h]
\centering
  \includegraphics[width=1.0\linewidth]{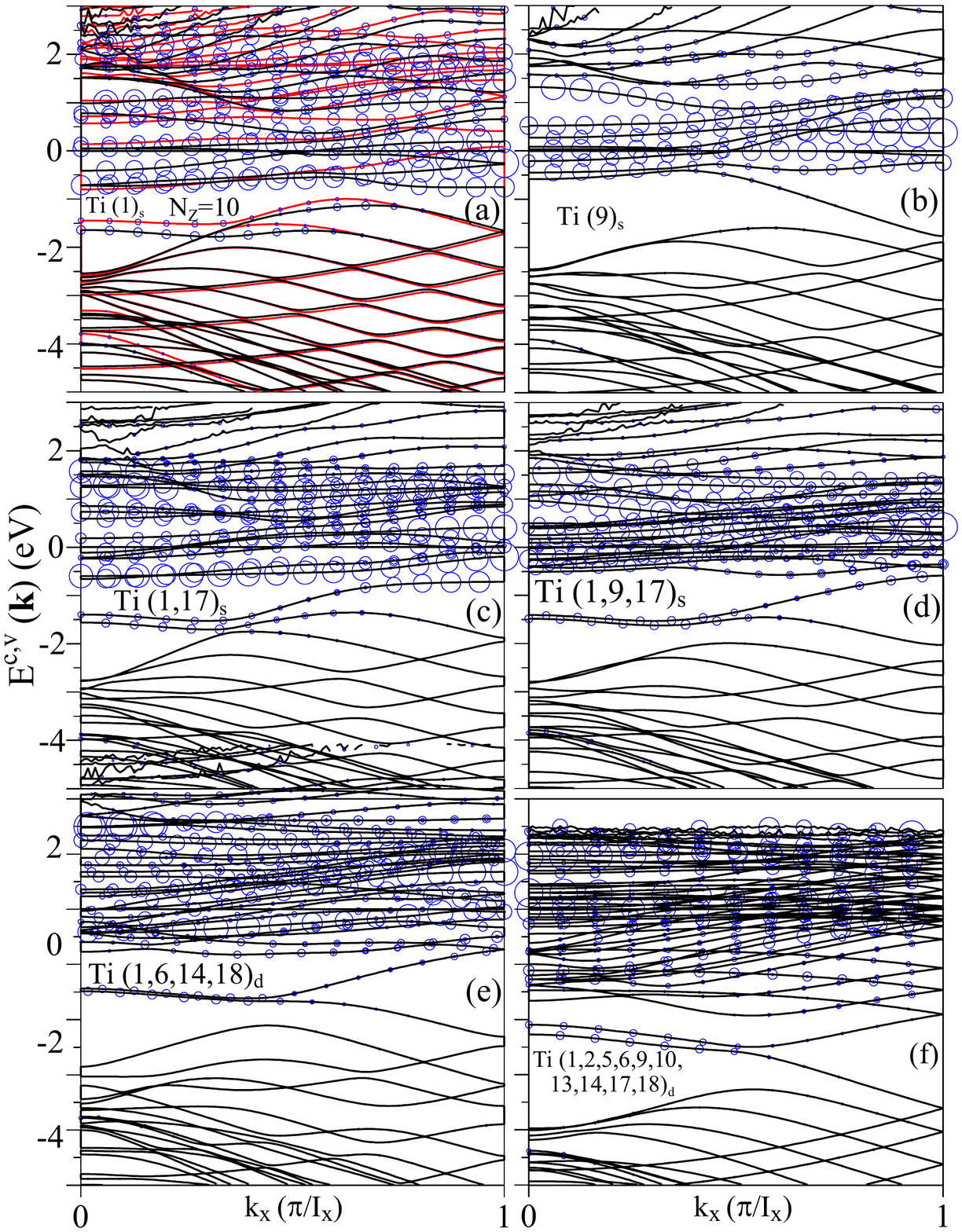}
  \caption{Similar plot as Fig. 2, but displayed for the ${N_z=10}$ zigzag systems under the distinct cases: (a) (1)$_s$, (b) (9)$_s$, (c) (1,17)$_d$, (d) (1,9,17)$_s$, (e) (1,6,14,18)$_d$ and (f) (1,2,5,6,9,10,13,14,17,18)$_s$.}
  \label{fgr:1}
\end{figure}

The Ti chemisorptions could induce the diversified band structures because of the edge structures. There are certain important differences between the zigzag and armchair systems.  As for the former, the partially flat bands across the Fermi level only survive under the single-adatom central adsorption (Fig. 9(b)). However, most of chemisorption cases, being shown in Figs. 9(a) and (c)-(f), lead to the dramatic changes in the low-lying energy bands because of the significant Ti-edge-C bondings; that is, the edge-C-dominated partially flat energy  dispersions are thoroughly absent. The zigzag systems might be FM, AFM and NM metals, as clearly revealed in Figs. 9(a)-9(f). in which the latter two need to be further examined from the spatial spin distributions. Apparently, this is purely due to the strong competition of edge carbon
atoms and Ti adatoms. The first kind of spin configuration, which is characterized by the distinct occupied states in the spin-up and spin-down energy bands, corresponds to the non-symmetric adatom distributions (Fig. 9(a)). Furthermore, the second and third kinds are closely associated with the symmetric ones (Figs. 9(b)-9(f)). They exhibit the spin-degenerate band structures..

\begin{figure}[h]
\centering
  \includegraphics[width=1.0\linewidth]{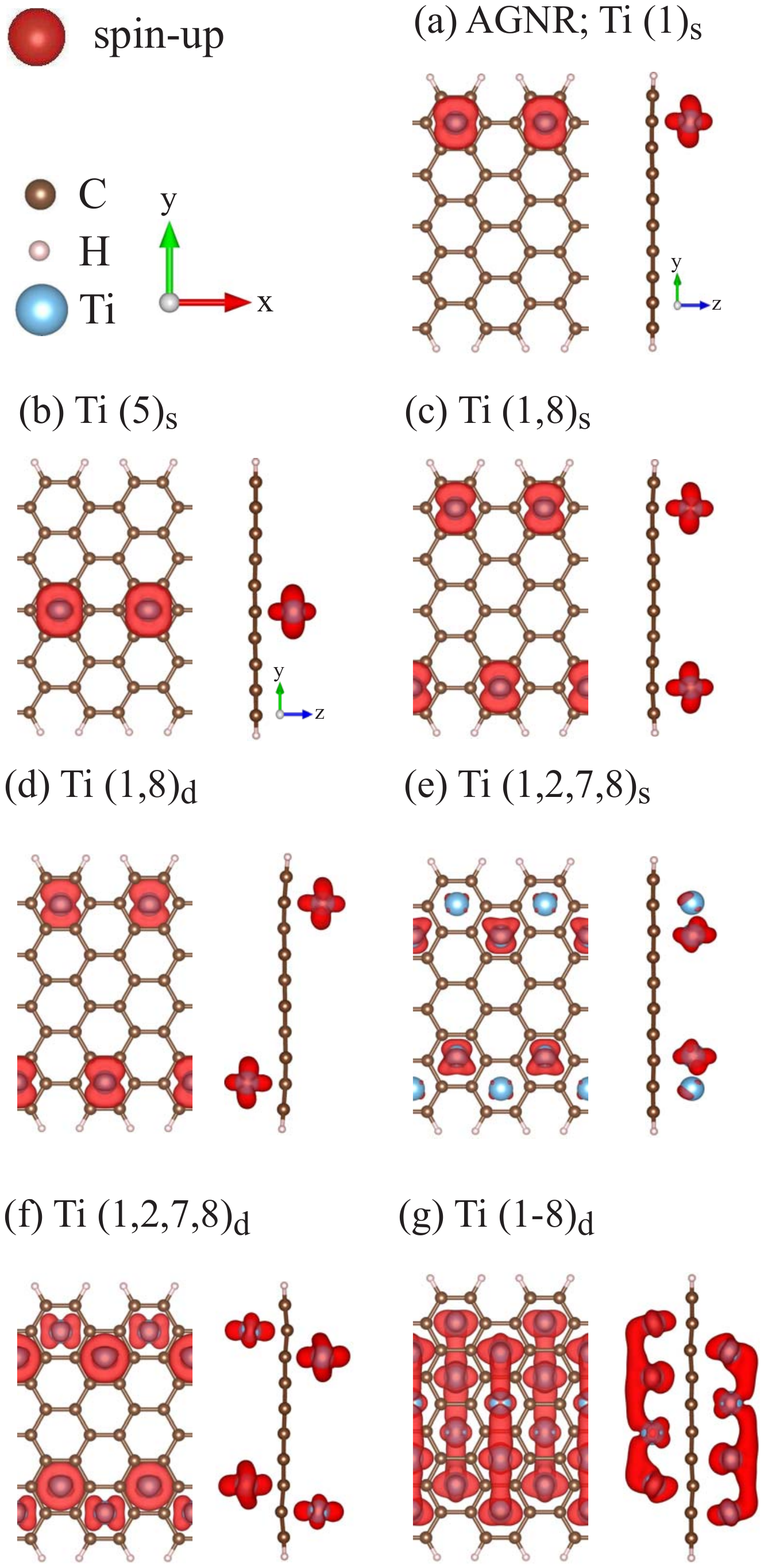}
  \caption{The spatial spin distributions for the Ti-adsorbed ${N_A=10}$ armchair systems arising from the various chemisorptions: (a) (1)$_s$, (b) (5)$_s$,  (c) (1,8)$_s$, (d)  (1,8)$_s$, (e) (1,2,7,8)$_s$, (f) (1,2,7,8)$_d$  and  (g) (1-8)$_d$. They/their insets are shown on the ${x-y}$/${y-z}$ plane.}
  \label{fgr:1}
\end{figure}

There exist the diverse various FM and AFM spin distributins. Any armchair systems have the distinct FM configuratons, being sensitive to the single- and double-side absorptions and concentrations. The single-adatom chemisorptions, e.g., (1)$_s$ and (5)$_s$ (Figs. 10(a) and Fig. 10(b)), could create the comparable magnetic moments (${\sim\,1.1-1.2} \mu B$ in Table 6) and slightly induce the identical spin arrangement of the neighboring carbon atoms. In general, the strength of magnetic response is proportional to the Ti concentration under the specific double-side adsorptions. For example, the red spin-related volumes grow with the inceeasing concentrations, (1)$_s$, (1,8)$_d$, (1,2,7,8)$_d$; (1-8)$_d$ respectively, in Figs. 10(a), 10(d), 10(f); 10(g), and so do the net magnetic moments. However, for the single-side chemisorptions, the neighboring  Ti adatoms might compete  with each other in spin states and thus create the abnormal  spin arrangements, e.g., the (1,2,7,8)$_s$ case with the smaller magnetic moment in Fig. 10(e).

\begin{figure}[h]
\centering
  \includegraphics[width=1.0\linewidth]{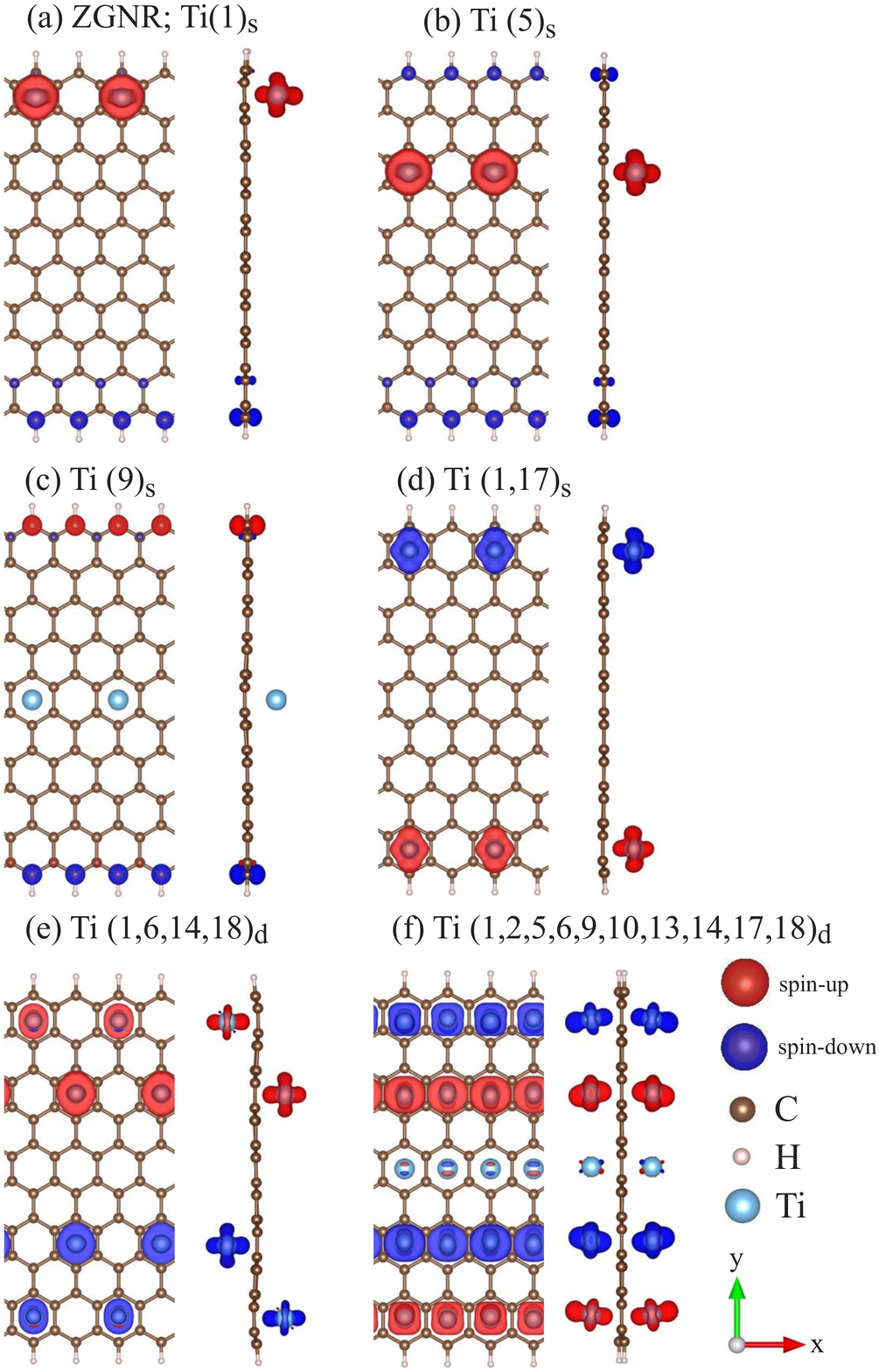}
  \caption{Similar plot as Fig, 4, but shown for the ${N_z=10}$ zigzag systems under the different chemisorptions: (a) (1)$_s$, (b) (5)$_s$, (c) (9)$_s$, (d) (1,17)$_s$, (e) (1,6,14,18)$_d$ and (f) (1,2,5,6,9,10,13,14,17,18)$_d$.}
  \label{fgr:1}
\end{figure}

The unusual competitions between zigzag carbons and Ti adatoms could lead to the diverse spin distributions, being strongly associated with the partial or full spin suppressions of the former and the latter. The FM and AFM configurations, respectively, correspond to the asymmetric and symmetric adatom distributions in zigzag systems. The single-adatom chemisorptions create the FM configuration ((1)$_s$ $\&$ (5)$_s$ in Figs. 11(a) and 11(b)) except for the symmetric case ((9)$_s$ in Fig. 11(c)). The guest adatom, which is located at (near) the zigzag boundary, has a wide spin-up distribution (red balls) and full destroys the spin-up-dominated arrangement from the neighboring carbons (even changes into the spin-down-determined configuration). Furthermore, there exists the spin-down dominant distribution on the other boundary (blue balls).
However, the spin state of the Ti adsorbate is absent under the symmetric adatom distribution, in which the pristine AFM configuration across the ribbon center is reduced, but remains similar. The absence of magnetism of Ti adatom might arise from the symmetric chemical and magnetic environments. If two/four Ti adatoms are very close to the distinct zigzag edges (Figs. 11(d)-11(f)), the AFM configurations purely due to the adsorbates are created and the spin states of the edge carbons are destroyed. In general, the spin arrangement near the specific boundary is FM for most of Ti adsorptions. It changes into the AFM configuration for very high concentrations, e.g., (1,2,5,6,9,10,13,14,17,18)$_d$ (Fig. 11(f)), where the middle adatoms do not induce the spin states. That is, the same edge or the distinct two edges display the AFM configurations simultaneously.

\begin{figure}[h]
\centering
  \includegraphics[width=1.0\linewidth]{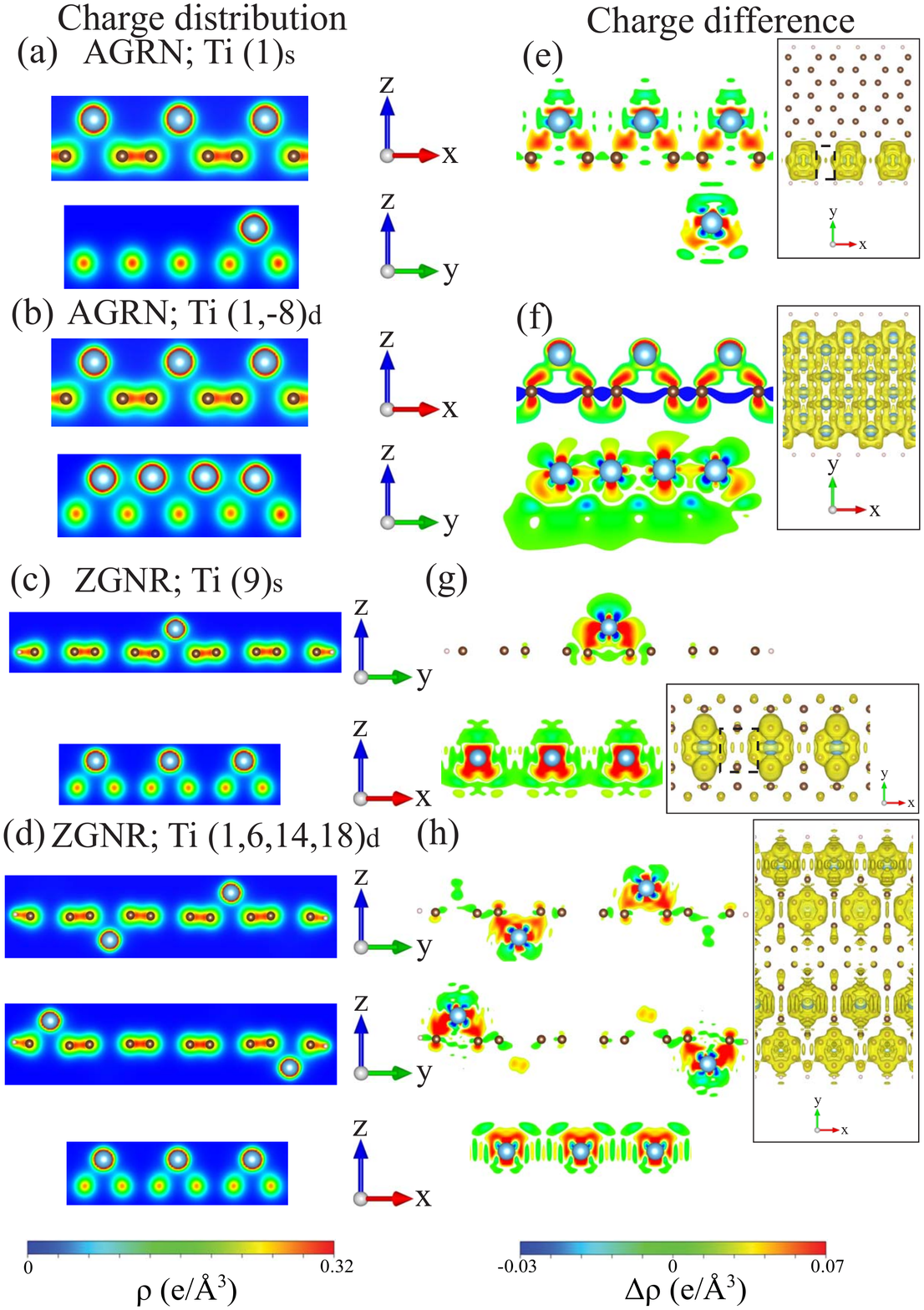}
  \caption{The spatial charge distributions and the differences after chemisorptions for the Ti-adsorbed systems. $\rho$/${\Delta\,\rho}$ on the $x$-$z$ plane for ${N_A=10}$ with Ti adatoms at (a)/(e) (1)$_s$ and (b)/(f) (1-8)$_d$. Also shown in the insets are ${\Delta\,\rho}$'s on the ${x-y}$ plane. Similar plots displayed for ${N_z=10}$ under Ti-adatom positions: (c)/(g) (9)$_s$ and (d)/(h) (1,6,14,18)$_d$.}
  \label{fgr:1}
\end{figure}

The transition metal adatoms have five kinds of $d$ orbitals, so that the multi-orbtial hybridizations will be clearly shown on the ${x-z}$, ${y-z}$ and ${x-y}$ planes. The chemical bonding between Ti and C is obvious and significant even for the single-adatom cases, regardless of the positions, such as (1)$_s$ and (9)$_s$ for ${N_A=10}$ and ${N_z=10}$, respectively (Figs. 12(a) and 12(c)). Its strength grows with the increasing adatom concentration e.g., ${(1-8)_d}$ and ${(1,6,14,18)_d}$ (Figs. 12(c) and 12(d)). The $\pi$ bonding of carbon ${2p_z}$ orbitals is distorted after chemisorptions, while it is still extended along the longitudinal and transverse directions. The carbon-dominated conduction bands crossing the Fermi level also make certain important contributions to the high free carriers. For any absorption cases, the large charge transfers exist between Ti guest adatoms and carbon host atoms on the $x$-$z$ and $y$-$z$planes, as clearly revealed in Figs. 12(e)-12(h). It can only identify the significant orbital hybridizations of (${3d_{xz},3d_{yz},3d_{xy},3d_{z^2},3d_{x^2-y^2}}$) five orbitals and ${2p_z}$ orbital. As to the Ti-Ti chemical bondings, the charge distributions are easily observed only on the ${y-z}$ plane at high concentration of armchair systems, e.g., ${(1-8)_d}$ in Fig. 12(b). The charge differences become obvious for any chemisorptions on the ${x-z}$ plane and/or the ${y-z}$ plane; furthermore, there are charge extensions and even overlaps on the ${x-y}$ plane at the optimal heights Ti adatoms (the insets in Figs. 12(e)-12(h)). The observable five-orbital hybridizations in Ti-Ti bonds are responsible for the low-lying Ti-dominated energy bands, being also one of the critical factors in the creation of the conduction electron density/the metallic behavior. In addition, it is very difficult to examine in detail about which ${3d}$ orbitals will make most of contributions to orbital interactions in Ti-C bonds under the current numerical calculations.

\begin{figure}[h]
\centering
  \includegraphics[width=1.0\linewidth]{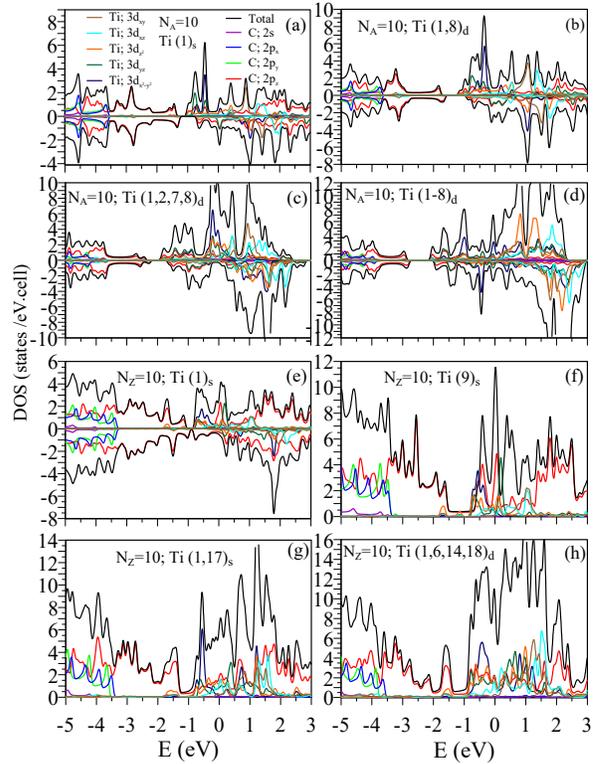}
  \caption{The orbital- and spin-decomposed DOSs of the ${N_A=10}$ Ti-adsorbed  systems under the various chemisorptions: (a) (1)$_s$, (b) (1,8)$_d$, (c) (1,2,7,8)$_d$; (d) (1,2,3,4,5,6,7,8)$_d$, and those for the ${N_z=10}$ ones with the adatom positions: (e) (1)$_s$, (f) (9)$_s$, (g) (1,17)$_d$, $\&$ (h) (1,6,14,18)$_d$.}
  \label{fgr:1}
\end{figure}

The orbital- and spin-projected DOSs in Ti-adsorbed graphene nanoribbons, as clearly shown in Figs. 13(a)-13(h), could provide the very complicated van Hove singularities and thus identify the significant multi-orbital hybridizations in Ti-C and Ti-Ti bonds. For any chemisorptions, there is an obvious DOS at the Fermi level, directly indicating the creation of the high free carrier density. This is closely related to carbon guest atoms and Ti quest adatom. The spin-split DOSs are revealed  in most of absorption cases (Figs. 13(a)-13(e)) except for the symmetric adatom distributions in zigzag systems (Figs. 13(f)-13(h)). The results further illustrate the Ti-induced spin states and their strong competitions with edge-carbon magnetic arrangement by the Ti-C chemical bondings. As to various orbital contributions to DOSs, ${2s}$, ${2p_x}$ and ${2p_y}$ of carbons (purple, blue and green curves) appear at ${E\le -3.4}$ eV even for the highest atadom concentration ((1-8)$_d$ in Fig. 13(d)). Such evidence strongly suggests their interactions with the outer five 3$d$ orbitals of Ti adatoms should be weak. However, the ${2p_z}$ orbitals experience rather high hybridizations with the ${3d_{xy}}$, ${3d_{xz}}$, ${3d_{yz}}$, ${3d_{z^2}}$ and ${3d_{x^2-y^2}}$, since the special structures of DOSs are seriously merged together in the range of ${-1}$ eV${\le E \le 3}$ eV/${-2}$ eV${\le E \le 3}$ eV for the dilute/heavy adsorptions (Figs. 13(a)-13(b) $\&$ Figs. 13(e)-13(h)/Figs. 13(c) and 13(d)). The Ti-Ti bonding is displayed in the enhancement of the metallic ${3d}$-band width. For example, energy band widths in the single-adatom and highest concentrations are, respectively, $\sim$1.5 eV and 3.0 eV (Figs. 13(a) and 13(d)).

\subsection{Bi}

\begin{figure}[h]
\centering
  \includegraphics[width=1.0\linewidth]{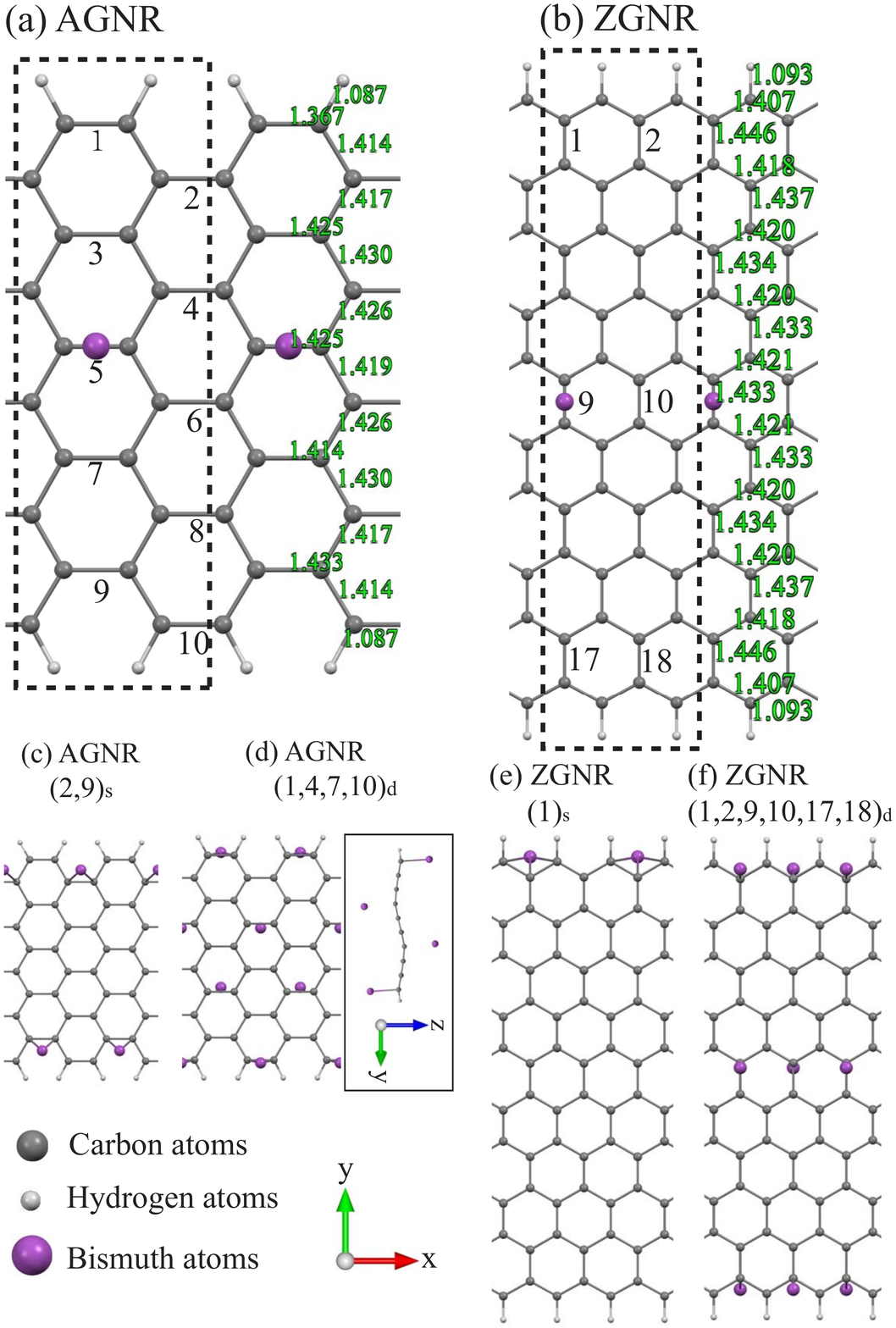}
  \caption{The optimal geometric structures of the Bi-adsorbed  ${N_A=10}$ armchair graphene nanoribbons with the initial adatom positions: (a) (5)$_s$ (c) (2,9)$_s$ $\&$ (d) (1,4,7,10)$_d$; those of  the ${N_Z=10}$ systems with the similar positions: (b) (9)$_s$, (e) (1)$_s$ $\&$ (f) (1,2,9,10,17,18)$_d$.}
  \label{fgr:1}
\end{figure}

Bismuth adatom chemisorption on graphene nanoribbon surface can create the unusual geometric structure, being sensitive to the position and concentration. A single Bi adatom near the ribbon center covers (5)$_s$ and (9)$_s$ adsorptions in the  ${N_A=10}$ and ${N_z=10}$ systems, respectively, as shown in Figs. 14(a) and 14(b). The optimal position is located at the bridge site and its height is about 3.85 $\AA$ (Table 6), as abserved in 2D Bi-adsorbed graphene [Rs]. When this adatom is close to the armchair/zigzag edge (Figs. 14(c) and 14(e)) , it shifts to the hollow site (along the ${\hat y}$ direction) about 0.3-0.4 $\AA$ (Table 6). Apparently, the height is reduced to the range of 2.4-2.5 $\AA$. The similar results are foumd in the two-atom edge cases, e.g., the Bi heights of (1,10)$_s$ and (2,9)$_s$ adsorptions (Fig. 14(c)). It is very difficult to form the high-concentration absorption system, since the Bi adatoms are very large and they have the significantly repulsive interactions within the sufficiently short distance. The maximum concentration is numerically examined to be about four and six Bi adatoms in the  ${N_A=10}$ and ${N_z=10}$ systems, respectively (Figs. 14(d) and 14(f)). The former even leads to the nonplanar/wave-like armchair graphene nanoribbon, indicating the quite strong Bi-C chemical bondings. Furthermore, there exist the abnormal heights in the peak and trough positions (Table 6), as shown by the ${y-z}$ plane projectionins for the (1,4,7,10)$_d$ adsorption of ${N_A=10}$  (inset in Fig. 14(d)). It is relatively difficult to observe the buckling structure in zigzag systems because of the larger widths. Most of the optimal structures show the planar geometries, or few of them display the wave-like ones accompanied with the bridge sites. As a result, the $\sigma$ bondngs of carbon atoms are hardly affected by Bi chemisorptions. The above-mentioned significant features clearly indicate the competition/cooperation among Bi-C, C-C and C-H bonds.

\begin{figure}[h]
\centering
  \includegraphics[width=1.0\linewidth]{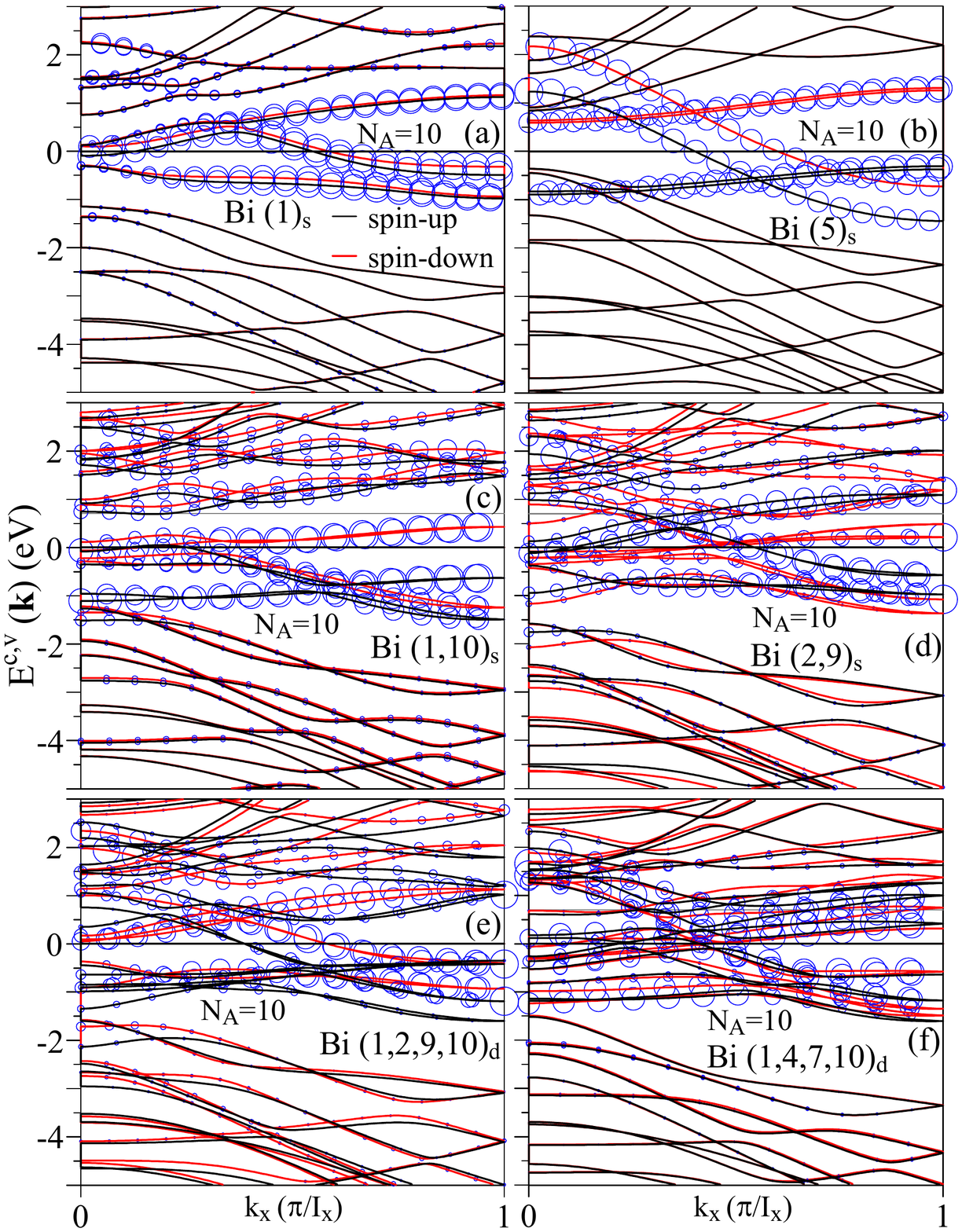}
  \caption{Energy bands for Bi-adsorbed ${N_A=10}$ armchair graphene nanoribbons under the distinct adsorptions: (a) (1)$_s$, (b) (5)$_s$, (c)  (1,10)$_s$, (d) (2,9)$_d$, (e) (1,2,9,10)$_d$  and  (f) (1,4,7,10)$_d$.}
  \label{fgr:1}
\end{figure}

All the Bi-adsorbed graphene nonoribbons are 1D metals with free carriers arising from the adatom chemisorptions, as clearly revealed in Figs. 15 and 16. The adatom-dominated energy bands occur near the Fermi level, in which they belong to the low-lying valence and conduction bands. Roughly speaking, they come to exist in the range of the original energy gap, especially for the central and dilute adsorptions. Such bands will have wider energy widths in the increase of adatom concentration. Most of them might have the weak energy dispersions with the high DOSs. The $\sigma$ bands (${ \sim E^v \le -3}$) due to carbon atoms are almost independent of Bi adtoms except for energy shifts, clearly reflecting the optimal geometric structures. However, the significant (C,Bi)-co-dominated band structure could occur at low energy. Moreover, whether band structures exhibit the spin-split behaviors depends the edge structure and the adatom distribution.

All the armchair Bi-adsorbed systems possess the spin-split electronic states, especially for the adatom-dominated energy bands. In general, the spin splittings are not obvious in the carbon-dominated energy bands. The occupied states near the Fermi level are not equivalent in the spin-up and spin-down ones, clearly indicating the net magnetic moment in each system. This is in great sharp with the non-magnetic behavior of the Al chemisorptions (the non-spin-split energy bands in Fig. 2). The magnetic properties are almost fully determined by the Bi adatoms in the armchair graphene nanoribbons (Fig. 15; discussed in Fig. 17). The net magnetic momentum is very sensitive to the adatom configuration (Table 1), and the maximum value could reach 1.35 ${\mu_B}$ per unit cell in the specific (1,2,9,10)$_d$ adsorption. As to the 1D conduction electrons, they are closely related to the Bi-dependent energy bands crossing the Fermi level. Compared with one free electron per adatom in alkali chemisorptions (Chap. 9.1), Bi adatom could induce more free carriers only under the single-adatom chemisorption, while the opposite is true for other adsorption cases.

\begin{figure}[h]
\centering
  \includegraphics[width=1.0\linewidth]{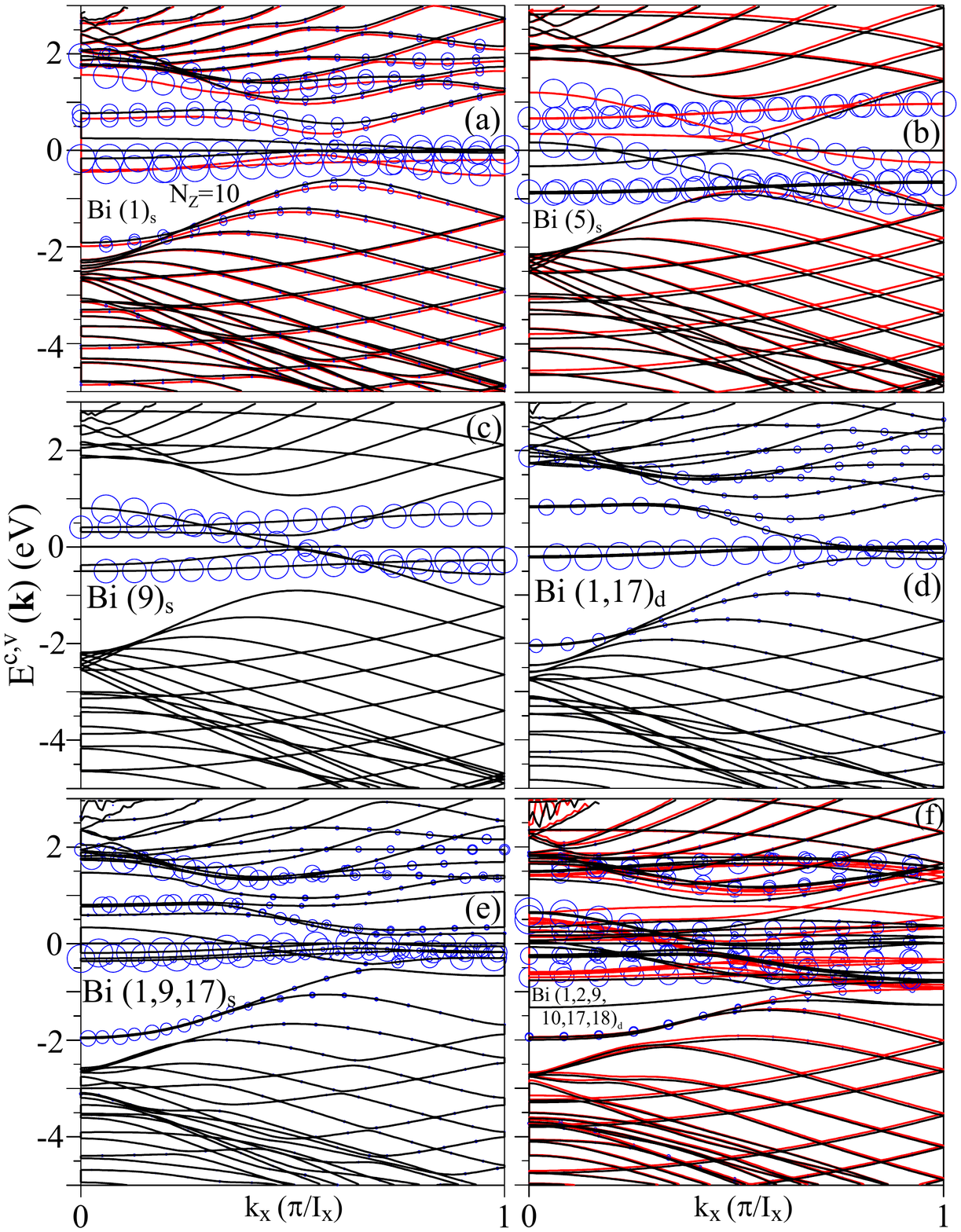}
  \caption{Similar plot as Fig. 2, but shown for the ${N_Z=10}$ zigzag systems under the various cases: (a) (1)$_s$, (b) (5)$_s$, (c) (9)$_s$, (d) (1,17)$_s$, (e) (1,9,17)$_s$ and  (f) (1,2,9,10,17,18)$_d$.}
  \label{fgr:1}
\end{figure}

The low-energy electronic structures are greatly diversified by the edge structure, and the zigzag and armchair Bi-adsorbed systems are very different from each other (Figs. 15 and 16). The edge-carbon- and Bi-induced energy bands in the zigzag nanoribbons appear near the Fermi level simultaneously, if the adatom is situated near the ribbon center (e.g., the (9)$_s$ case in Fig. 16(c)). The former remain a pair of partially flat dispersions crossing the Fermi level (without blue circles), without the spin splitting. Furthermore, the latter exhibits the unusual behavior in the absence of spin-split electronic states; that is, it does not have any magnetism. This unique magnetic property will be thoroughly discussed in the spatial spin distribution (Fig. 18(c)). The low-lying energy bands will become very complicated for the adatoms close to the zigzag boundaries, such as the (1)$_s$ and (5)$_s$ adsorptions in Figs. 16(a) and 16(b), respectively. Specifically, the (1)$_s$ case exhibits the drastic changes in the originally edge-localized bands, and it has the strongly hybridized energy bands in the range of ${-1}$ eV${\le\,E^{c,v}\le\,1}$ eV  in the presence of significant contributions from edge carbons and bismuths simultaneously. These clearly indicate the important chemical bondings of host and quest atoms. The similar results could be found in the higher-concentration edge adsorptions, e.g., the ${(1,17)_s}$, ${(1,9,17)_s}$ and ${(1,2,9,10,17,18)_d}$ chemisorptions in Figs. 16(d), 16(e) and 16(f), respectively. As to the magnetic properties, most of the Bi-adsorbed zigzag  nanoribbons display the spin splittings and the FM configuration (Table 1; spin density in Fig. 18). It is very easy to measure the magnetic response, since the significant magnetic moment is about ${0.5-1.6}$ ${\mu_{B}}$ per unit cell. Only few of absorption configurations possess the spin-degenerate energy bands: one-adatom center (Fig. 16(c)), two-adatom  zigzag edges (Fig. 16(d)), and three-adatom center $\&$ edges (Fig. 16(e)). Apparently, the net moment is vanishing and the spin arrangement belongs to the AFM configuration.

\begin{figure}[h]
\centering
  \includegraphics[width=1.0\linewidth]{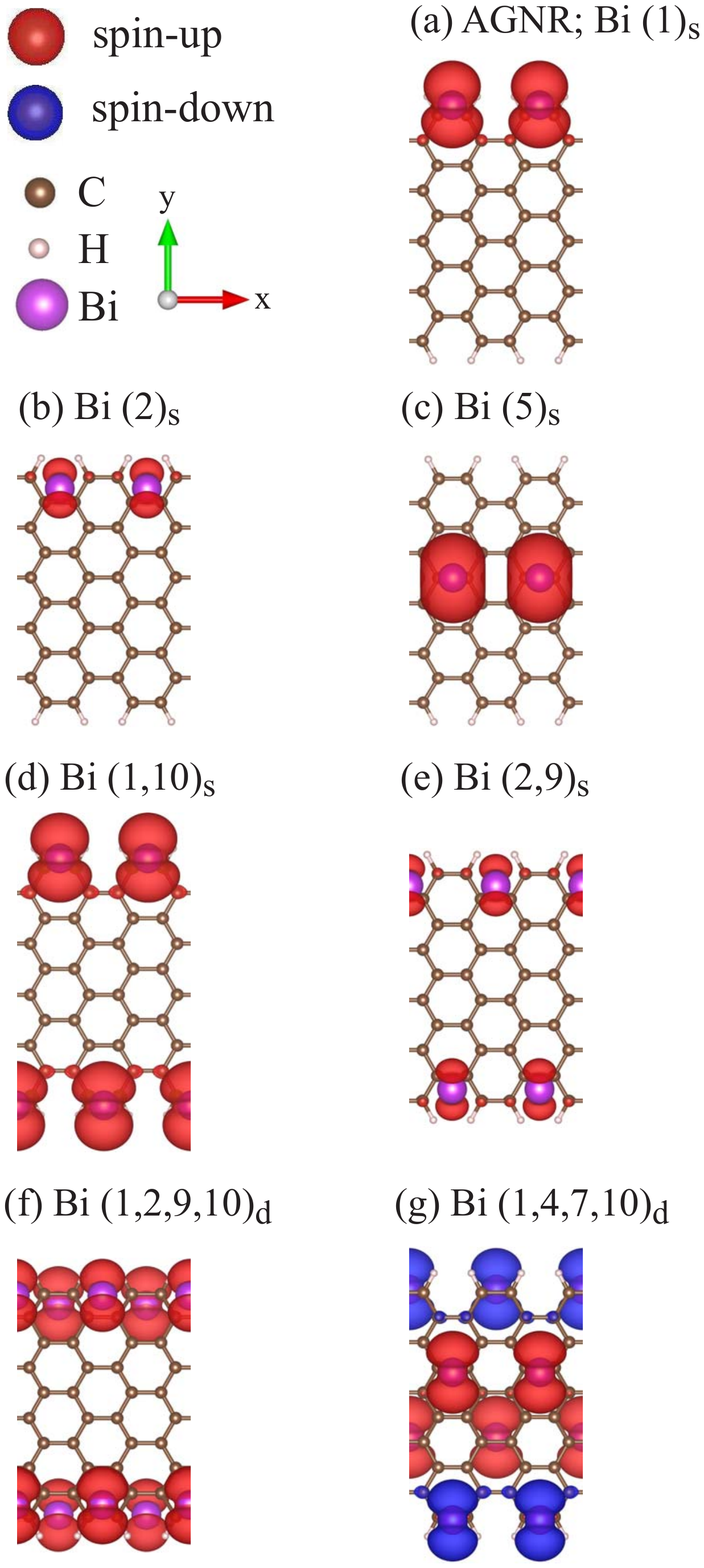}
  \caption{The spatial spin distributions of the Bi-adsorbed ${N_A=10}$ armchair systems due to the various adsorptions: (a) (1)$_s$, (b) (2)$_s$,  (c) (5)$_s$, (d)  (1,10)$_s$, (e) (2,9)$_d$, (f) (1,2,9,10)$_d$  and  (g) (1,4,7,10)$_d$. They are shown on the ${x-y}$ plane.}
  \label{fgr:1}
\end{figure}

The spin configuration (Figs. 17 and 18) and the strength of magnetic response (Table 1) are greatly diversified by the Bi chemisorptions on the graphene nanoribbon surfaces. The isolated bismuth atoms could themselves induce the intrinsic magnetic mement, so only the FM and AFM spin distributions come to exist after the quest-adatom adsorptions. All the armchair systems show the FM configurations, and most/few of the zigzag ones exhibit the FM/AFM spin arrangement. For the former, the Bi guest atoms create  the spin-up distribution, also leading to the minor and similar magnetic configuration of the nieghboring carbon atoms (Figs. 17(a)-17(g)). From the single-adatom cases, the spatial spin densities are strongly dependent on its position, e.g., the obvious differences among the (1)$_s$, (2)$_s$ and (5)$_s$ chemisorptions (Figs. 17(a)-17(c); 0.56 ${\mu_B}$, 0.11 ${\mu_B}$ and 1.18 ${\mu_B}$ in Table 1). This is related to the chemical environment experienced by the Bi adatoms. The magnetic responses are also very sensitive to the adatom concentration, in which, a simple linear relation between them is absent. For example, the net magnetic moment is large under the ${(1,10)_s}$ and ${(1,2,9,10)_d}$ adsorptions (Figs. 17(d) and 17(f); 1.05 ${\mu_B}$ and 1.35 ${\mu_B}$), while it is small for the ${(2,9)_s}$ and ${(1,4,7,10)_d}$ adsorptions (Figs. 17(e) and 17(g); 0.21 ${\mu_B}$). The magnetic properties are also affected by the terminated C-H bonds, as implied from two-adatom cases. Four-adatom chemisorptions show that the neighboring Bi adatoms could enhance the strong magnetic response. However, the edge and near-center adatoms, respectively, create the spin-down and spin-up distributions and thus induce the weaker magnetic response.

\begin{figure}[h]
\centering
  \includegraphics[width=1.0\linewidth]{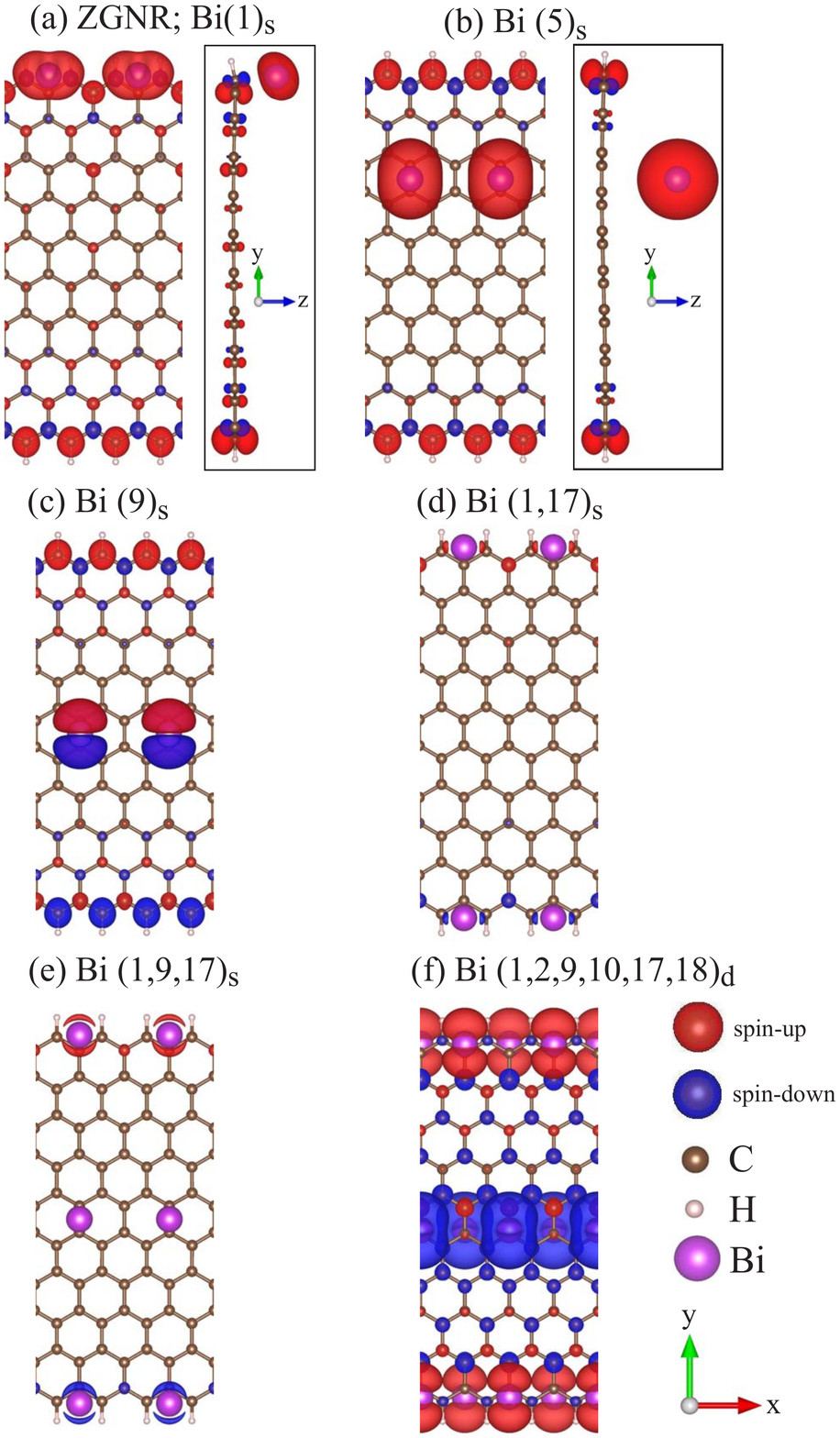}
  \caption{Similar plot as Fig, 4, but displayed for the ${N_Z=10}$ zigzag systems for the different cases: (a) (1)$_s$, (b) (5)$_s$, (c) (9)$_s$, (d) (1,17)$_s$, (e) (1,9,17)$_s$  and  (f) (1,2,9,10,17,18)$_d$. Also shown in in insets of (a) and (b) are those on the ${y-z}$ plane.}
  \label{fgr:1}
\end{figure}

On the other side, the zigzag systems have the quite strong competition/cooperation between Bi adatoms and edge carbons in the spin state/orbital hybridization, leading to the diverse and unique magnetic phenomena. When the Bi-adsorbed system remains a symmetric geometry under the single adsorptions, the adatom simultaneously exhibits the spin-up and spin-down magnetic moments with the same magnitude ((9)$_s$ in Fig. 18(c)), or the Bi-dependent magnetism is thoroughly vanishing. Also, the spin arrangement of edge carbons keeps the AFM configuration across the ribbon center. However, the Bi adatom in an asymmetric  edge distribution itself only generates the spin-up state, e.g., (1)$_s$ and (5)$_s$ (Figs. 18(a) and 18(b)). It has a strong effect on the edge-carbon spins and even alters the magnetic dominance in the zigzag edges. The obvious FM configurations are revealed along the longitudinal and transverse directions. Specifically, the (5)$_s$ adsorption shows the maximum magnetic moment (${1.59 \mu_B}$ in Table 6).

\begin{figure}[h]
\centering
  \includegraphics[width=1.0\linewidth]{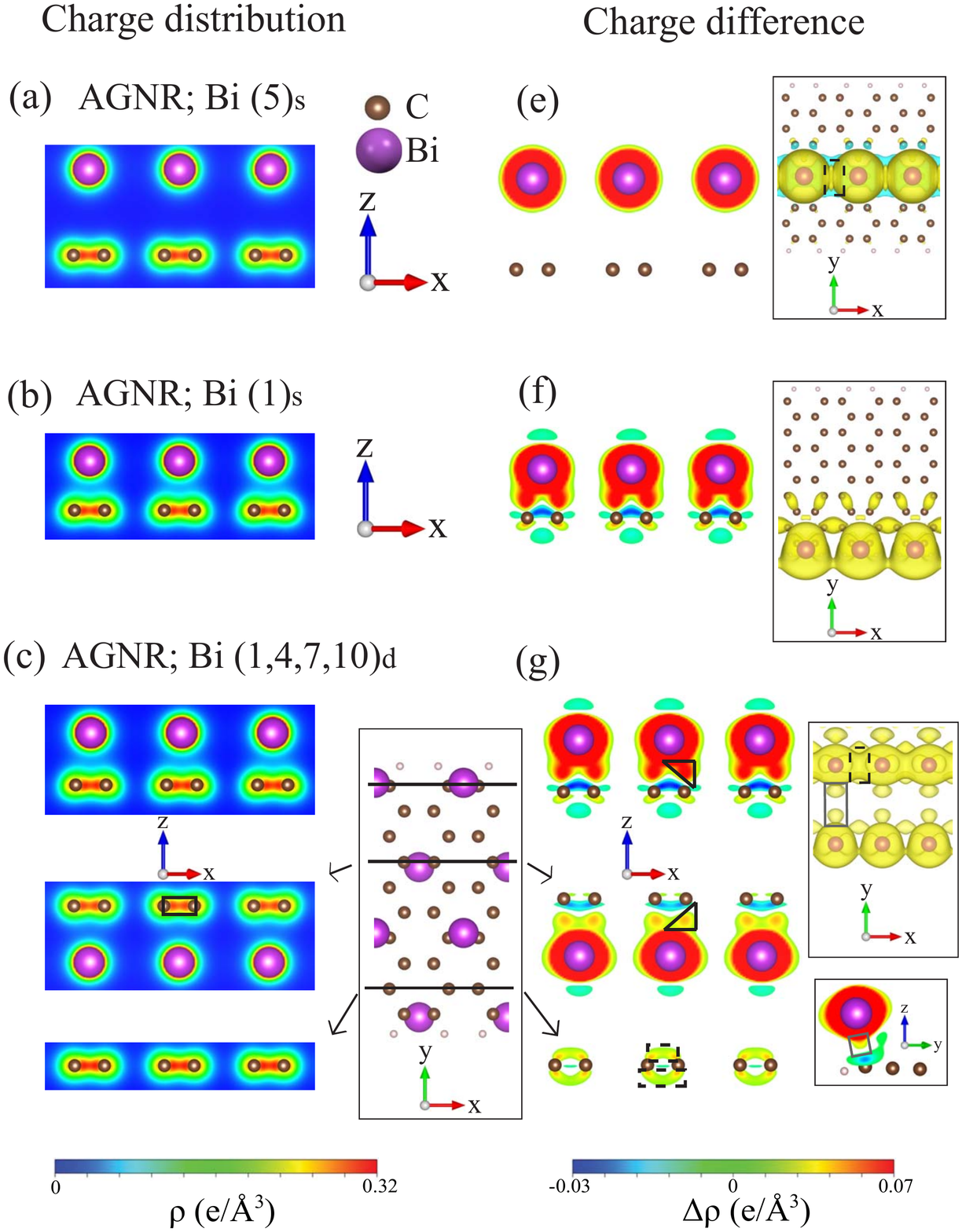}
  \caption{The spatial charge distributions and the differences after chemisorptions. $\rho$'s on the ${x-z}$ plane for the ${N_A=10}$ systems with Bi adatoms at (a) (5)$_s$, (b) (1)$_s$ $\&$ (1,4,7,10)$_d$, respectively, correspond to ${\Delta\,\rho}$'s in (d), (e) $\&$ (f). The latter are also indicated on the ${x-y}$ plane in the insets.}
  \label{fgr:1}
\end{figure}

\begin{figure}[h]
\centering
  \includegraphics[width=1.0\linewidth]{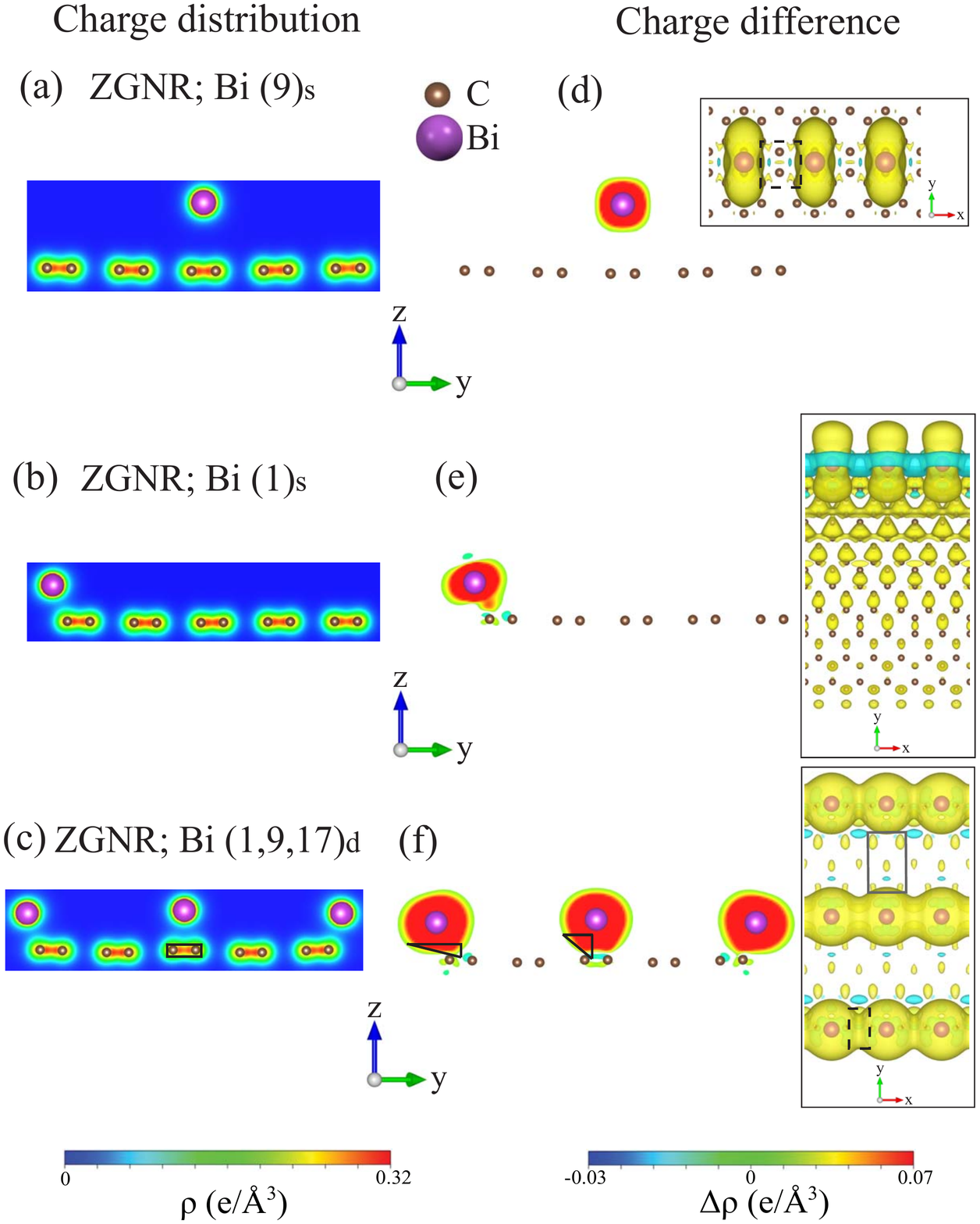}
  \caption{Same plot as Fig. 19, but shown for the ${N_Z=10}$ systems under.
$\rho$'s on the ${y-z}$ plane.}
  \label{fgr:1}
\end{figure}

The group-V elements have three valence electrons in the outmost orbitals, they can form the complex multi-orbital hybridizations after chemisorptions on graphene nanoribbon surfaces. In general, the Bi adatoms possess the ${(6p_x,6p_y,6p_z)}$ orbitals which will take part in the Bi-C and Bi-Bi bonds. The multi-orbital hybridizations are clearly revealed in the spatial charge distributions, as observed in Fig. 19. The single-adatom central adsorptions are shown for armchair and zigzag graphene nanoribbons in Figs. 19(a)/19(d) and Figs. 20(a)/20(d), respectively, in which ${\rho}$ and ${\Delta \rho}$ almost have no charge distributions between guest and host atoms. This suggests the rather weak interactions in Bi-C bonds under the very high optimal position (${3.85-3.89}$ $\AA$ in Table 6). However, figs. 19(d) and 20(d) in ${x-z}$ and ${x-y}$ planes clearly show that charge variation is obvious near the Bi adatoms, indicating the significant Bi-Bi bonds. The main orbital hybridizations cover ${6p_x-6p_x}$, ${6p_y-6p_y}$ and ${6p_z-6p_z}$, e.g., the first one in the armchair system (the black dashed rectangle in the inset of Fig. 19(d)). On the other hand, for most of Bi-adatom adsorptions (the others), there exist the observable/apparent charge distributions in Bi-C bonds and more complicated orbital interactions in Bi-Bi bonds. That is, the asymmetric single-adatom and higher-concentrations adsorptions could create more orbital hybridizations in Bi-C and Bi-Bi bonds, such as, (1)$_s$ and (1,4,7,10)$_d$ in armchair systems (Figs. 19(b) and 19(c)); (1)$_s$ and (1,9,17)$_d$ in zigzag cases (Fig. 20(b) and 103(c)). The significant Bi-C bondings have induced the charge distributions between them, accompanied with the partial distortions of the $\pi$ bondings. The distorted and extended  $\pi$ bondings and the  Bi-Bi bonds dominate the metallic behavior, especially for the latter. This is consistent with the orbital-projected DOSs (discussed in Fig. 19). The important orbital hybridizations in Bi-C bonds, as observed from the charge differences, cover ${6p_z-2p_z}$ (clearly indicated in Fig. 19(e)/Fig. 19(f)/Fig. 20(e)/Fig 20(f)), ${6p_z-2p_y}$ (Fig. 20(e)/Fig. 20(f)), and ${6p_z-2p_x}$ (Fig. 19(e)/Fig. 19(f)).The first kind has the strongest orbital hybridization among them.

\begin{figure}[h]
\centering
  \includegraphics[width=1.0\linewidth]{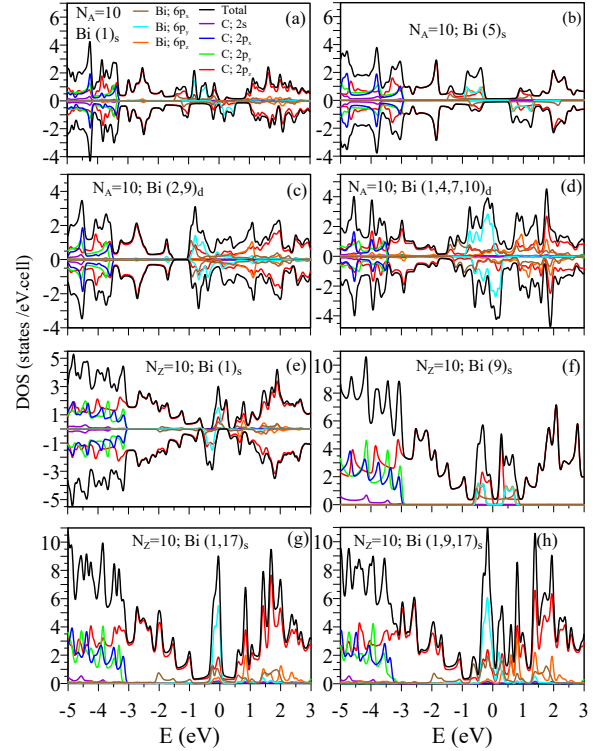}
  \caption{The orbital- and spin-decomposed DOSs for the ${N_A=10}$ armchair systems under the various cases: (a) (1)$_s$, (b) (5)$_s$, (c) (2,9)$_d$; (d) (1,4,7,10)$_d$, and those for the ${N_Z=10}$ systems  with the adatom adsorptions: (a) (1)$_s$, (b) (9)$_s$, (c) (1,17)$_s$, $\&$ (d) (1,9,17)$_s$.v}
  \label{fgr:1}
\end{figure}

The orbital- and spin-decomposed DOSs obviously reveal the metallic behavior, the magnetic configuration, and the concise orbital hybridizations in chemical bonds. The Bi adatom chemisorptions can induce a finite DOS at the Fermi level (Figs. 21(a)-21(h)), in which its value directly reflects the energy dispersions crossing ${E_F=0}$ (Figs. 15 and 16). Most of cases exhibit  very high DOSs (Figs. 21(c)-21(h)), accompanied with the weakly dispersive bands. However, the dilute absorptions in armchair systems might display low DOSs ((1)$_s$ and (5)$_s$, respectively, in Figs. 21(a) and 21(b)) because of the highly dispersive bands. The spin-up and spin-down DOSs are different in all the Bi-adsorbed graphene nanoribbons except for the symmetric adatom distributions in zigzag systems (Fig. 21(f)-21(h)). Their difference in the covered area means the FM configuration and is proportional to the net magnetic moment. For example, it is about half for the ratio between the (1)$_s$ and (5)$_s$ armchair cases, and so does the net magnetic moment (0.56 $\mu_B$ and 1.18 $\mu_B$ in Table 6). The AFM and NM configurations cannot be distinguished from the spin-degenerate DOS, while they are confirmed by the spatial spin distributions (Fig. 18(c)-18(e)). As to the orbital hybridizations, the DOSs, which arise from the ${(2s,2p_x,2p_y)}$ orbitals of carbon atoms, occur at ${E\le\,3}$ eV and are hardly affected by the surface adsorptions. That is, the planar $\sigma$ bonding almost keeps the similar, being consistent with the optimal geometric structures.  Only the ${2p_z}$ orbitals might take part in the Bi-C bonding. In general, their contributions to the total DOS in the range of ${-1.5}$ eV${\le E \le 2.5}$ eV show the special structures merged those arising from the Bi ${(6p_x,6p_y,6p_z)}$ orbitals. The ${6p_z-2p_z}$ bondings (the orange and red curves) are strongest among three kinds of orbital hybridizations in Bi-C bonds, as clearly indicated from the strength and number of combined van Hove singularities. However, the emergence is greatly reduced in the single-adatom central adsorptions ((5)$_s$ and (9)$_s$, respectively, in armchair and zigzag systems. This is responsible for the very weak Bi-C bonding. As to the significant hybridizations of ${(6p_x,6p_y,6p_z)}$ in Bi-Bi bandings, they are obviously revealed in the three-orbital-dependent merged structures, in which the intensities are enhanced in the increase of adatom concentration. The 6p$_x$-6p$_x$ and 6p$_y$-6p$_y$ orbital hybridizations are significant and dominating, as observed from the orbital-dependent van Hove singularities.

\section{Concluding remarks}

The metal atadoms, Al/Ti/Bi, can induce the metallic behaviors, being in sharp contrast with the alkali ones. They clearly display the important differences in the essential properties. The Al and Ti guest atoms exhibit the hollow-site optimal positions, while the former might have the $y$-direction shifts, especially for the non-symmetric distributions. The similar shifts are revealed in the deviated  bridge-site Bi adatoms. The adatom chemisorptions are relatively easily observed in the Ti chemisorptions with the largest binding energies; that is, the Ti-adatom adsorption has the highest concentration. Energy bands, which crossing the Fermi level, mainly arise from carbon atoms (metal atoms) for Al-adsorbed systems (Ti- and Al-doped ones). There exist the Al-dominated valence bands at ${E^v\sim\-3}$ eV and the partial adatom contributions to the conduction bands, being consistent with the ${3p_z-2p_z}$ and ${(3p_x+3p_y)-2p_z}$ orbital hybridizations in Al-C bonds. This feature is also supported by the spatial charge distributions/DOSs. At higher adatom concentrations, the significant ${3p_x-3p_x}$ $\&$ ${3p_y-3p_y}$ hybridizations in Al-Al bonds are indicated by the enhanced Al-dependent conduction bands/DOSs with the wider energy  widths. The multi-orbital hybridizations in Ti-C bonds are very complicated, in which five orbitals of adatom, ($d_{xy}$, $d_{xz}$, $d_{yz}$, $d_{z^2}$, and $d_{x^2-y^2}$ have the strong interactions with ${2p_z}$ orbitals of carbons.
Such orbitals also take part in the Ti-Ti chemical bondings. It is very difficult to observe more complex orbital interactions in the other condensed-matter systems. Three kinds of observable Bi-C bondings, ${6p_z-2p_z}$, ${6p_x-2p_z}$ $\&$ ${6p_y-2p_z}$, are characterized by the charge distributions/DOSs except for the symmetric single-adatom adsorptions. The important ${(6p_x,6p_y,6p_z)-(6p_x,6p_y,6p_z)}$ orbital hybridizations in Bi-Bi bonds  are responsible for the Bi-induced low-lying energy bands. As to the magnetic properties, the Al adatoms do not create the spin distributions, but their interactions with the zigzag carbon atoms can destroy the latter's ones and thus create the FM or NM configurations. Armchair and zigzag Al-adsorbed graphene naoribbons, respectively, belong to the NM and AFM/FM/NM metals. On the other side, most of Ti- and Bi-adsorbed asymmetric systems exhibit the FM configurations under the themself-induced spin states and the greatly reduced edge-carbon magnetic moments. Furthermore, the symmetric adatom distributions in zigzag systems might lead to the very complicated AFM configurations, being in sharp contrast with the pristine magnetic configuration. The AFM spin arrangements is strongly associated with the edge carbons and adsorbates. For the Ti adatoms, the strength of magnetic response only relies on the concentration for the double-side adsorptions.

\vskip 0.6 truecm
\par\noindent {\bf ACKNOWLEDGMENTS}
\vskip 0.3 truecm

This work was supported by the Physics Division, National Center for Theoretical Sciences (South), the Nation Science Council of Taiwan (Grant No. NSC 102-2112-M-006-007-MY3).

\footnotesize{
\bibliography{achemso} 

\providecommand*{\mcitethebibliography}{\thebibliography}
\csname @ifundefined\endcsname{endmcitethebibliography}
{\let\endmcitethebibliography\endthebibliography}{}
\begin{mcitethebibliography}{33}
\providecommand*{\natexlab}[1]{#1}
\providecommand*{\mciteSetBstSublistMode}[1]{}
\providecommand*{\mciteSetBstMaxWidthForm}[2]{}
\providecommand*{\mciteBstWouldAddEndPuncttrue}
  {\def\EndOfBibitem{\unskip.}}
\providecommand*{\mciteBstWouldAddEndPunctfalse}
  {\let\EndOfBibitem\relax}
\providecommand*{\mciteSetBstMidEndSepPunct}[3]{}
\providecommand*{\mciteSetBstSublistLabelBeginEnd}[3]{}
\providecommand*{\EndOfBibitem}{}
\mciteSetBstSublistMode{f}
\mciteSetBstMaxWidthForm{subitem}
{(\emph{\alph{mcitesubitemcount}})}
\mciteSetBstSublistLabelBeginEnd{\mcitemaxwidthsubitemform\space}
{\relax}{\relax}

\bibitem[Lin \emph{et~al.}(2015)Lin, Gong, Lu, Wu, Wang, Guan, Angell, Chen,
  Yang, and Hwang]{lin2015ultrafast}
M.-C. Lin, M.~Gong, B.~Lu, Y.~Wu, D.-Y. Wang, M.~Guan, M.~Angell, C.~Chen,
  J.~Yang and B.-J. Hwang, \emph{Nature}, 2015, \textbf{520}, 324\relax
\mciteBstWouldAddEndPuncttrue
\mciteSetBstMidEndSepPunct{\mcitedefaultmidpunct}
{\mcitedefaultendpunct}{\mcitedefaultseppunct}\relax
\EndOfBibitem
\bibitem[Elia \emph{et~al.}(2016)Elia, Marquardt, Hoeppner, Fantini, Lin,
  Knipping, Peters, Drillet, Passerini, and Hahn]{elia2016overview}
G.~A. Elia, K.~Marquardt, K.~Hoeppner, S.~Fantini, R.~Lin, E.~Knipping,
  W.~Peters, J.-F. Drillet, S.~Passerini and R.~Hahn, \emph{Advanced
  Materials}, 2016, \textbf{28}, 7564--7579\relax
\mciteBstWouldAddEndPuncttrue
\mciteSetBstMidEndSepPunct{\mcitedefaultmidpunct}
{\mcitedefaultendpunct}{\mcitedefaultseppunct}\relax
\EndOfBibitem
\bibitem[Ao \emph{et~al.}(2008)Ao, Yang, Li, and Jiang]{ao2008enhancement}
Z.~Ao, J.~Yang, S.~Li and Q.~Jiang, \emph{Chemical Physics Letters}, 2008,
  \textbf{461}, 276--279\relax
\mciteBstWouldAddEndPuncttrue
\mciteSetBstMidEndSepPunct{\mcitedefaultmidpunct}
{\mcitedefaultendpunct}{\mcitedefaultseppunct}\relax
\EndOfBibitem
\bibitem[Chi and Zhao(2009)]{chi2009adsorption}
M.~Chi and Y.-P. Zhao, \emph{Computational Materials Science}, 2009,
  \textbf{46}, 1085--1090\relax
\mciteBstWouldAddEndPuncttrue
\mciteSetBstMidEndSepPunct{\mcitedefaultmidpunct}
{\mcitedefaultendpunct}{\mcitedefaultseppunct}\relax
\EndOfBibitem
\bibitem[Ao \emph{et~al.}(2009)Ao, Jiang, Zhang, Tan, and Li]{ao2009doped}
Z.~Ao, Q.~Jiang, R.~Zhang, T.~Tan and S.~Li, \emph{Journal of Applied Physics},
  2009, \textbf{105}, 074307\relax
\mciteBstWouldAddEndPuncttrue
\mciteSetBstMidEndSepPunct{\mcitedefaultmidpunct}
{\mcitedefaultendpunct}{\mcitedefaultseppunct}\relax
\EndOfBibitem
\bibitem[Lin \emph{et~al.}(2017)Lin, Lin, Tran, Su, and Lin]{lin2017feature}
S.-Y. Lin, Y.-T. Lin, N.~T.~T. Tran, W.-P. Su and M.-F. Lin, \emph{Carbon},
  2017, \textbf{120}, 209--218\relax
\mciteBstWouldAddEndPuncttrue
\mciteSetBstMidEndSepPunct{\mcitedefaultmidpunct}
{\mcitedefaultendpunct}{\mcitedefaultseppunct}\relax
\EndOfBibitem
\bibitem[Chen \emph{et~al.}(2016)Chen, Huang, Convertino, Coletti, Chang, Shiu,
  Cheng, Lin, Heun, and Chien]{chen2016efficient}
J.-W. Chen, H.-C. Huang, D.~Convertino, C.~Coletti, L.-Y. Chang, H.-W. Shiu,
  C.-M. Cheng, M.-F. Lin, S.~Heun and F.~S.-S. Chien, \emph{Carbon}, 2016,
  \textbf{109}, 300--305\relax
\mciteBstWouldAddEndPuncttrue
\mciteSetBstMidEndSepPunct{\mcitedefaultmidpunct}
{\mcitedefaultendpunct}{\mcitedefaultseppunct}\relax
\EndOfBibitem
\bibitem[Chan \emph{et~al.}(2008)Chan, Neaton, and Cohen]{chan2008first}
K.~T. Chan, J.~Neaton and M.~L. Cohen, \emph{Physical Review B}, 2008,
  \textbf{77}, 235430\relax
\mciteBstWouldAddEndPuncttrue
\mciteSetBstMidEndSepPunct{\mcitedefaultmidpunct}
{\mcitedefaultendpunct}{\mcitedefaultseppunct}\relax
\EndOfBibitem
\bibitem[Mashoff \emph{et~al.}(2013)Mashoff, Takamura, Tanabe, Hibino, Beltram,
  and Heun]{mashoff2013hydrogen}
T.~Mashoff, M.~Takamura, S.~Tanabe, H.~Hibino, F.~Beltram and S.~Heun,
  \emph{Applied Physics Letters}, 2013, \textbf{103}, 013903\relax
\mciteBstWouldAddEndPuncttrue
\mciteSetBstMidEndSepPunct{\mcitedefaultmidpunct}
{\mcitedefaultendpunct}{\mcitedefaultseppunct}\relax
\EndOfBibitem
\bibitem[Hofmann(2006)]{hofmann2006surfaces}
P.~Hofmann, \emph{Progress in surface science}, 2006, \textbf{81},
  191--245\relax
\mciteBstWouldAddEndPuncttrue
\mciteSetBstMidEndSepPunct{\mcitedefaultmidpunct}
{\mcitedefaultendpunct}{\mcitedefaultseppunct}\relax
\EndOfBibitem
\bibitem[Li \emph{et~al.}(2008)Li, Checkelsky, Hor, Uher, Hebard, Cava, and
  Ong]{li2008phase}
L.~Li, J.~G. Checkelsky, Y.~S. Hor, C.~Uher, A.~F. Hebard, R.~J. Cava and
  N.~Ong, \emph{Science}, 2008, \textbf{321}, 547--550\relax
\mciteBstWouldAddEndPuncttrue
\mciteSetBstMidEndSepPunct{\mcitedefaultmidpunct}
{\mcitedefaultendpunct}{\mcitedefaultseppunct}\relax
\EndOfBibitem
\bibitem[Bobaru \emph{et~al.}(2012)Bobaru, Gaudry, De~Weerd, Ledieu, and
  Fourn{\'e}e]{bobaru2012competing}
S.~Bobaru, {\'E}.~Gaudry, M.-C. De~Weerd, J.~Ledieu and V.~Fourn{\'e}e,
  \emph{Physical Review B}, 2012, \textbf{86}, 214201\relax
\mciteBstWouldAddEndPuncttrue
\mciteSetBstMidEndSepPunct{\mcitedefaultmidpunct}
{\mcitedefaultendpunct}{\mcitedefaultseppunct}\relax
\EndOfBibitem
\bibitem[Hirahara \emph{et~al.}(2006)Hirahara, Nagao, Matsuda, Bihlmayer,
  Chulkov, Koroteev, Echenique, Saito, and Hasegawa]{hirahara2006role}
T.~Hirahara, T.~Nagao, I.~Matsuda, G.~Bihlmayer, E.~Chulkov, Y.~M. Koroteev,
  P.~Echenique, M.~Saito and S.~Hasegawa, \emph{Physical review letters}, 2006,
  \textbf{97}, 146803\relax
\mciteBstWouldAddEndPuncttrue
\mciteSetBstMidEndSepPunct{\mcitedefaultmidpunct}
{\mcitedefaultendpunct}{\mcitedefaultseppunct}\relax
\EndOfBibitem
\bibitem[Hirahara \emph{et~al.}(2012)Hirahara, Fukui, Shirasawa, Yamada,
  Aitani, Miyazaki, Matsunami, Kimura, Takahashi, and
  Hasegawa]{hirahara2012atomic}
T.~Hirahara, N.~Fukui, T.~Shirasawa, M.~Yamada, M.~Aitani, H.~Miyazaki,
  M.~Matsunami, S.~Kimura, T.~Takahashi and S.~Hasegawa, \emph{Physical review
  letters}, 2012, \textbf{109}, 227401\relax
\mciteBstWouldAddEndPuncttrue
\mciteSetBstMidEndSepPunct{\mcitedefaultmidpunct}
{\mcitedefaultendpunct}{\mcitedefaultseppunct}\relax
\EndOfBibitem
\bibitem[Yang \emph{et~al.}(2012)Yang, Miao, Wang, Yao, Zhu, Song, Wang, Xu,
  Fedorov, and Sun]{yang2012spatial}
F.~Yang, L.~Miao, Z.~Wang, M.-Y. Yao, F.~Zhu, Y.~Song, M.-X. Wang, J.-P. Xu,
  A.~V. Fedorov and Z.~Sun, \emph{Physical review letters}, 2012, \textbf{109},
  016801\relax
\mciteBstWouldAddEndPuncttrue
\mciteSetBstMidEndSepPunct{\mcitedefaultmidpunct}
{\mcitedefaultendpunct}{\mcitedefaultseppunct}\relax
\EndOfBibitem
\bibitem[Wang \emph{et~al.}(2013)Wang, Yao, Ming, Miao, Zhu, Liu, Gao, Qian,
  Jia, and Liu]{wang2013creation}
Z.~Wang, M.-Y. Yao, W.~Ming, L.~Miao, F.~Zhu, C.~Liu, C.~Gao, D.~Qian, J.-F.
  Jia and F.~Liu, \emph{Nature communications}, 2013, \textbf{4}, 1384\relax
\mciteBstWouldAddEndPuncttrue
\mciteSetBstMidEndSepPunct{\mcitedefaultmidpunct}
{\mcitedefaultendpunct}{\mcitedefaultseppunct}\relax
\EndOfBibitem
\bibitem[Hirahara \emph{et~al.}(2015)Hirahara, Shirai, Hajiri, Matsunami,
  Tanaka, Kimura, Hasegawa, and Kobayashi]{hirahara2015role}
T.~Hirahara, T.~Shirai, T.~Hajiri, M.~Matsunami, K.~Tanaka, S.~Kimura,
  S.~Hasegawa and K.~Kobayashi, \emph{Physical review letters}, 2015,
  \textbf{115}, 106803\relax
\mciteBstWouldAddEndPuncttrue
\mciteSetBstMidEndSepPunct{\mcitedefaultmidpunct}
{\mcitedefaultendpunct}{\mcitedefaultseppunct}\relax
\EndOfBibitem
\bibitem[Sabater \emph{et~al.}(2013)Sabater, Gos{\'a}lbez-Mart{\'\i}nez,
  Fern{\'a}ndez-Rossier, Rodrigo, Untiedt, and
  Palacios]{sabater2013topologically}
C.~Sabater, D.~Gos{\'a}lbez-Mart{\'\i}nez, J.~Fern{\'a}ndez-Rossier,
  J.~Rodrigo, C.~Untiedt and J.~Palacios, \emph{Physical review letters}, 2013,
  \textbf{110}, 176802\relax
\mciteBstWouldAddEndPuncttrue
\mciteSetBstMidEndSepPunct{\mcitedefaultmidpunct}
{\mcitedefaultendpunct}{\mcitedefaultseppunct}\relax
\EndOfBibitem
\bibitem[Chen \emph{et~al.}(2017)Chen, Wu, and Lin]{chen2017novel}
S.-C. Chen, J.-Y. Wu and M.-F. Lin, \emph{arXiv preprint arXiv:1709.03289},
  2017\relax
\mciteBstWouldAddEndPuncttrue
\mciteSetBstMidEndSepPunct{\mcitedefaultmidpunct}
{\mcitedefaultendpunct}{\mcitedefaultseppunct}\relax
\EndOfBibitem
\bibitem[Black \emph{et~al.}(2002)Black, Lin, Cronin, Rabin, and
  Dresselhaus]{black2002infrared}
M.~Black, Y.-M. Lin, S.~Cronin, O.~Rabin and M.~Dresselhaus, \emph{Physical
  Review B}, 2002, \textbf{65}, 195417\relax
\mciteBstWouldAddEndPuncttrue
\mciteSetBstMidEndSepPunct{\mcitedefaultmidpunct}
{\mcitedefaultendpunct}{\mcitedefaultseppunct}\relax
\EndOfBibitem
\bibitem[Liu and Allen(1995)]{PhysRevB.52.1566}
Y.~Liu and R.~E. Allen, \emph{Phys. Rev. B}, 1995, \textbf{52},
  1566--1577\relax
\mciteBstWouldAddEndPuncttrue
\mciteSetBstMidEndSepPunct{\mcitedefaultmidpunct}
{\mcitedefaultendpunct}{\mcitedefaultseppunct}\relax
\EndOfBibitem
\bibitem[Jason~Lee(2015)]{Jason2015Two}
W.-L. W. D.-X.~Y. Jason~Lee, Wen-Chuan~Tian, \emph{Scientific Reports}, 2015,
  \textbf{5}, year\relax
\mciteBstWouldAddEndPuncttrue
\mciteSetBstMidEndSepPunct{\mcitedefaultmidpunct}
{\mcitedefaultendpunct}{\mcitedefaultseppunct}\relax
\EndOfBibitem
\bibitem[Wanekaya(2011)]{wanekaya2011applications}
A.~K. Wanekaya, \emph{Analyst}, 2011, \textbf{136}, 4383--4391\relax
\mciteBstWouldAddEndPuncttrue
\mciteSetBstMidEndSepPunct{\mcitedefaultmidpunct}
{\mcitedefaultendpunct}{\mcitedefaultseppunct}\relax
\EndOfBibitem
\bibitem[Shan \emph{et~al.}(2009)Shan, Zhang, Xue, Zhang, Cosnier, and
  Ding]{shan2009polycrystalline}
D.~Shan, J.~Zhang, H.-G. Xue, Y.-C. Zhang, S.~Cosnier and S.-N. Ding,
  \emph{Biosensors and Bioelectronics}, 2009, \textbf{24}, 3671--3676\relax
\mciteBstWouldAddEndPuncttrue
\mciteSetBstMidEndSepPunct{\mcitedefaultmidpunct}
{\mcitedefaultendpunct}{\mcitedefaultseppunct}\relax
\EndOfBibitem
\bibitem[Li \emph{et~al.}(2013)Li, Trujillo, Fu, Patterson, Fei, Xu, Deng,
  Smirnov, and Luo]{li2013bismuth}
Y.~Li, M.~A. Trujillo, E.~Fu, B.~Patterson, L.~Fei, Y.~Xu, S.~Deng, S.~Smirnov
  and H.~Luo, \emph{Journal of Materials Chemistry A}, 2013, \textbf{1},
  12123--12127\relax
\mciteBstWouldAddEndPuncttrue
\mciteSetBstMidEndSepPunct{\mcitedefaultmidpunct}
{\mcitedefaultendpunct}{\mcitedefaultseppunct}\relax
\EndOfBibitem
\bibitem[Chen \emph{et~al.}(2015)Chen, Su, Chang, Cheng, Chong, Huang, and
  Lin]{chen2015long}
H.-H. Chen, S.~Su, S.-L. Chang, B.-Y. Cheng, C.-W. Chong, J.~Huang and M.~F.
  Lin, \emph{Carbon}, 2015, \textbf{93}, 180--186\relax
\mciteBstWouldAddEndPuncttrue
\mciteSetBstMidEndSepPunct{\mcitedefaultmidpunct}
{\mcitedefaultendpunct}{\mcitedefaultseppunct}\relax
\EndOfBibitem
\bibitem[Chen \emph{et~al.}(2015)Chen, Su, Chang, Cheng, Chen, Chen, Lin, and
  Huang]{chen2015tailoring}
H.-H. Chen, S.~Su, S.-L. Chang, B.-Y. Cheng, S.~Chen, H.-Y. Chen, M.~F. Lin and
  J.~Huang, \emph{Scientific reports}, 2015, \textbf{5}, year\relax
\mciteBstWouldAddEndPuncttrue
\mciteSetBstMidEndSepPunct{\mcitedefaultmidpunct}
{\mcitedefaultendpunct}{\mcitedefaultseppunct}\relax
\EndOfBibitem
\bibitem[Lin \emph{et~al.}(2016)Lin, Chang, Chen, Su, Huang, and
  Lin]{lin2016substrate}
S.-Y. Lin, S.-L. Chang, H.-H. Chen, S.-H. Su, J.-C. Huang and M.-F. Lin,
  \emph{Physical Chemistry Chemical Physics}, 2016, \textbf{18},
  18978--18984\relax
\mciteBstWouldAddEndPuncttrue
\mciteSetBstMidEndSepPunct{\mcitedefaultmidpunct}
{\mcitedefaultendpunct}{\mcitedefaultseppunct}\relax
\EndOfBibitem
\bibitem[Akt{\"u}rk and Tomak(2010)]{akturk2010bismuth}
O.~{\"U}. Akt{\"u}rk and M.~Tomak, \emph{Applied Physics Letters}, 2010,
  \textbf{96}, 081914\relax
\mciteBstWouldAddEndPuncttrue
\mciteSetBstMidEndSepPunct{\mcitedefaultmidpunct}
{\mcitedefaultendpunct}{\mcitedefaultseppunct}\relax
\EndOfBibitem
\bibitem[Hsu \emph{et~al.}(2013)Hsu, Ozolins, and Chuang]{hsu2013first}
C.-H. Hsu, V.~Ozolins and F.-C. Chuang, \emph{Surface Science}, 2013,
  \textbf{616}, 149--154\relax
\mciteBstWouldAddEndPuncttrue
\mciteSetBstMidEndSepPunct{\mcitedefaultmidpunct}
{\mcitedefaultendpunct}{\mcitedefaultseppunct}\relax
\EndOfBibitem
\bibitem[Kresse and Furthm{\"u}ller(1996)]{kresse1996efficient}
G.~Kresse and J.~Furthm{\"u}ller, \emph{Physical review B}, 1996, \textbf{54},
  11169\relax
\mciteBstWouldAddEndPuncttrue
\mciteSetBstMidEndSepPunct{\mcitedefaultmidpunct}
{\mcitedefaultendpunct}{\mcitedefaultseppunct}\relax
\EndOfBibitem
\bibitem[Grimme(2006)]{grimme2006semiempirical}
S.~Grimme, \emph{Journal of computational chemistry}, 2006, \textbf{27},
  1787--1799\relax
\mciteBstWouldAddEndPuncttrue
\mciteSetBstMidEndSepPunct{\mcitedefaultmidpunct}
{\mcitedefaultendpunct}{\mcitedefaultseppunct}\relax
\EndOfBibitem
\bibitem[Perdew \emph{et~al.}(1996)Perdew, Burke, and
  Ernzerhof]{perdew1996generalized}
J.~P. Perdew, K.~Burke and M.~Ernzerhof, \emph{Physical review letters}, 1996,
  \textbf{77}, 3865\relax
\mciteBstWouldAddEndPuncttrue
\mciteSetBstMidEndSepPunct{\mcitedefaultmidpunct}
{\mcitedefaultendpunct}{\mcitedefaultseppunct}\relax
\EndOfBibitem
\end{mcitethebibliography}
\bibliographystyle{rsc} 
}

\begin{@twocolumnfalse}
\newpage
\begin{table}
\caption{}
\label{tbl:3}
\begin{center}
  \begin{tabular}{cccccc}
 & Eb (eV) & Height ({\AA}) & Adatom y-shift ({\AA})& M ($\mu$B) \\
\hline
\hline
AGNR; Al (1)$_s$ & -0.641 & 2.071 & 0.319 & 0 \\
  (5)$_s$ & -0.639 & 2.207 & x & 0 \\
  (1,8)$_s$ & -0.636 & 2.063 & 0.252 & 0 \\
  (1,8)$_d$ & -0.636 & 2.067 & 0.289 & 0 \\
  (2,7)$_s$ & -0.453 & 2.077 & x & 0 \\
  (2,7)$_d$ & -1.969 & 2.065 & x & 0 \\
  (1,2,7,8)$_d$ & -0.714 & 2.153 & 0.144; 1.231 & 0 \\
\hline
ZGNR; Al (1)$_s$ & -1.606 & 2.049 & 0.172 & 0.56/FM \\
  (3)$_s$ & -1.229 & 1.996 & 0.072 & 0.52/FM \\
  (5)$_s$ & -1.074 & 2.076 & 0.064 & 0.18/FM \\
  (9)$_s$ & -1.142 & 2.135 & x & 0/AFM \\
  (1,17)$_s$ & -1.789 & 1.999 & 0.317 & 0 \\
\hline
AGNR; Bi (1)$_s$ & -0.786 & 2.397 & 0.502 & 0.56/FM \\
  (2)$_s$ & -0.071 & 2.483 & 0.551 & 0.11/FM \\
  (5)$_s$ & -0.033 & 3.854 & x & 1.18/FM \\
  (1,10)$_s$ & -0.443 & 2.404 & 0.481 & 1.08/FM \\
  (2,9)$_s$ & -0.024 & 2.442 & 0.571 & 0.21/FM \\
  (1,2,9,10)$_d$ & -0.615 & 1.933/3.319 & 1.011; -0.529 & 1.35/FM \\
  (1,4,7,10)$_d$ & -0.276 & 2.521/3.046 & -0.235; 0.154 & 0.21/FM \\
\hline
ZGNR; Bi (1)$_s$ & -1.142 & 2.325 & 1.648 & 0.54/FM \\
  (5)$_s$ & -0.697 & 3.944 & x & 1.59/FM \\
  (9)$_s$ & -0.319 & 3.889 & x & 0/AFM \\
  (1,17)$_s$ & -1.233 & 2.260 & 1.653 & 0/AFM \\
  (1,9,17)$_s$ & -0.909 & 2.294/2.247 & 1.536/x & 0/AFM \\
\hline
AGNR; Ti (1)$_s$ & -2.828 & 1.777 & x & 1.28/FM \\
  (5)$_s$ & -2.591 & 1.784 & x & 1.52/FM \\
  (1,8)$_s$ & -2.790 & 1.769 & x & 2.27/FM \\
  (2,7)$_s$ & -2.542 & 1.838 & x & 2.91/FM \\
  (2,7)$_d$ & -2.545 & 1.811 & x & 2.89/FM \\
  (1,2,7,8)$_s$ & -3.783 & 1.729/1.859 & x & 1.46/FM \\
  (1,2,7,8)$_d$ & -2.854 & 1.506/2.109 & x & 4.22/FM \\
  (1,2,4,5,7,8)$_d$ & -3.376 & 1.779/1.919 & x & 4.16/FM \\
  (1,2,3,4,5,6,7,8)$_d$ & -3.454 & 1.822/1.985 & x & 5.93/FM \\
\hline
ZGNR; Ti (1)$_s$ & -2.774 & 1.711 & x & 1.46/FM \\
  (5)$_s$ & -2.273 & 1.728 & x & 1.52/FM \\
  (9)$_s$ & -1.991 & 1.629 & x & 0/AFM \\
  (1,17)$_s$ & -2.564 & 1.670 & x & 0/AFM \\
  (1,9,17)$_s$ & -2.207 & 1.618 & x & 0/AFM \\
  (1,6,14,18)$_d$ & -2.354 & 1.588/1.607 & x & 0/AFM \\
  (1,2,5,6,9,10,13,14,17,18)$_d$ & -2.606 & 1.665/1.718 & x & 0/AFM \\
\hline
\hline
\\
\end{tabular}
\end{center}
\end{table}
\end{@twocolumnfalse}


\end{document}